\documentclass[a4paper,fleqn,usenatbib]{mnras}

\usepackage{txfonts}

\usepackage[T1]{fontenc}
\usepackage{ae,aecompl}

\usepackage{psfig}   
\usepackage{graphicx}
\usepackage{amssymb}
\usepackage{subfigure}

\title[Spatial distribution of dense cores]{On the spatial distributions of dense cores in Orion B}

\author[R. J. Parker]{Richard  J. Parker\thanks{E-mail: R.Parker@sheffield.ac.uk}\thanks{Royal Society Dorothy Hodgkin Fellow} \vspace*{0.1cm}\\
Department of Physics and Astronomy, The University of Sheffield, Hicks Building, Hounsfield Road, Sheffield, S3 7RH, UK}

\begin{document}

                             
\pagerange{\pageref{firstpage}--\pageref{lastpage}} \pubyear{2018}

\maketitle

\label{firstpage}

\begin{abstract}
 We quantify the spatial distributions of dense cores in three spatially distinct areas of the Orion~B star-forming region. For L1622, NGC\,2068/NGC\,2071 and  NGC\,2023/NGC\,2024 we measure the amount of spatial substructure using the $\mathcal{Q}$-parameter and find all three regions to be spatially substructured ($\mathcal{Q} < 0.8$). We quantify the amount of mass segregation using $\Lambda_{\rm MSR}$ and find that the most massive cores are mildly mass segregated in NGC\,2068/NGC\,2071 ($\Lambda_{\rm MSR} \sim 2$), and very mass segregated in NGC\,2023/NGC\,2024 ($\Lambda_{\rm MSR} = 28^{+13}_{-10}$ for the four most massive cores). Whereas the most massive cores in L1622 are not in areas of relatively high surface density, or deeper gravitational potentials, the massive cores in NGC\,2068/NGC\,2071 and  NGC\,2023/NGC\,2024 are significantly so. Given the low density (10\,cores\,pc$^{-2}$) and spatial substructure of cores in Orion~B, the mass segregation cannot be dynamical. Our results are also inconsistent with simulations in which the most massive stars form via competitive accretion, and instead hint that magnetic fields may be important in influencing the primordial spatial distributions of gas and stars in star-forming regions.
\end{abstract}

\begin{keywords}
stars: formation -- massive -- kinematics and dynamics -- star clusters: general -- methods: numerical
\end{keywords}

\section{Introduction}

One of the great challenges in astrophysics is to understand the star formation process. Stars form in groups where the mean stellar density exceeds that of the Galactic field by several orders of magnitude \citep{Lada03,Porras03,Bressert10}. At these high densities, environmental conditions can affect the outcome of star formation due to early disc truncation and disruption \citep{Scally01,Adams04,Zwart16}, and the properties of primordial binary and multiple systems are rapidly altered due to internal and external dynamical evolution \citep{Kroupa95a,Reipurth14}.

Due to the rapid changes experienced by infant stars, it is imperative to quantify and understand the early stages of star formation, such as the initial distribution of dense cores that will eventually form one or more stars. Studies of the mass function of prestellar cores \citep{Andre10,Konyves10} have shown that they follow a similar distribution to the stellar initial mass function (IMF), but with the core mass function (CMF) shifted to higher masses. However, it is unclear if the stellar IMF is set by this core mass function (CMF), which is simply shifted due to lower masses by a star formation efficiency of $\sim 1/3$ \citep{Alves07}, or whether the form of the IMF is independent of the CMF \citep[see e.g.][for a review]{Offner14}.

In addition to the mass distribution of cores, a wealth of spatial and kinematic information now exists for these objects. The general spatio-kinematic picture is that cores form along dense filaments \citep[e.g.][]{Andre10,Hacar13,Henshaw16,Smith16,Kainulainen17}, with low (subvirial) velocity dispersions \citep[e.g.][]{Peretto06,Schneider10,Kauffmann13,Foster15}. However, it is unclear how much of a signature the stars that form from dense cores retain from the initial conditions of the gas. Several studies have pointed out similarities between the amount of spatial substructure in young stars and the interstellar medium \citep{Hoyle53,Elmegreen96,Elmegreen02,Gouliermis14}, although analysis of simulations suggest the stars and gas become decoupled early in the star formation process \citep[and similarities in their spatial distributions may be unrelated,][]{Bate05,Kruijssen12a,Parker15c}.

The spatial distribution of the most massive stars in star-forming regions has been the topic of numerous observational \citep{Hillenbrand98,Raboud98,deGrijs02,Littlefair04,Allison09a,Wright14,Kuhn17,Parker17a} and theoretical studies \citep{Bonnell98,Moeckel09a,Moeckel09b,Allison10,Olczak11,Maschberger11,Girichidis12,Parker14b,Kuznetsova15,Dominguez17}, with the goal of understanding if the formation channel of massive stars produces a different spatial distribution to that of low-mass stars -- so-called mass segregation. Initially, mass segregation was thought to be a natural outcome of the competitive accretion theory for star formation \citep{Zinnecker82,Bonnell98b,Bonnell01}, where the most massive stars would from in the most gas-rich regions of the cluster, which in turn would likely be the more central regions. However, extensive analysis of several hydrodynamic simulations of star formation \citep{Parker15a,Parker17b} suggest that competitive accretion does not necessarily lead to mass segregation, ostensibly because the star-forming region is substructured and the dense cores/stars have cannot fully interact with one another during the formation process.

Given that most star formation theories appear not to predict a different spatial distribution for the most massive stars, any observed variation as a function of stellar mass that could not be explained through dynamical processes \citep{McMillan07,Allison10,Parker14b}, or attributed to stochasticity in the star formation process, would require a new theoretical framework for star formation. So far, most studies have focused on the spatial distributions of pre-main sequence stars, but it is unclear if observed cores could be primordially mass segregated \citep[e.g.][]{Elmegreen14}. 

To fully address these issues, a comprehensive comparison between the spatial distributions of cores and stars in observations and simulations is required. Recently, \citet{Kirk16a} used SCUBA--2 data from the James Clerk Maxwell Telescope (JCMT) to identify prestellar and protostellar cores in the Orion~B star-forming region. Using the 850$\mu$m flux as a tracer or proxy for core mass, \citet{Kirk16b} quantified the spatial substructure of three spatially distinct areas of Orion~B; the Linds Dark Nebula 1622 (hereafter L1622) and the NGC\,2068/NGC\,2071 and NGC\,2023/NGC\,2024 regions.

\citet{Kirk16b} found that none of the three subregions are spatially substructured according the the $\mathcal{Q}$-parameter \citep{Cartwright04,Cartwright09b}, which is surprising as all three regions appear visually substructured. The authors also claim to find mass segregation of the cores, but using the group segregation ratio method \citep{Kirk10,Kirk14}. However, \citet{Parker15b} find serious flaws in this technique, to the extent that it may not accurately find or quantify mass segregation in spatially substructured star-forming regions. For these reasons, we have decided to revisit the JCMT SCUBA--2 data from \citet{Kirk16a,Kirk16b} to produce an independent analysis of the spatial distributions of the dense cores in Orion~B. 

In this paper, we use the same Orion~B data as \citet{Kirk16b} to quantify the spatial distribution of cores, but add two further diagnostics to the analysis; the $\Lambda_{\rm MSR}$ mass segregation ratio \citep{Allison09a} and the local gravitational potential difference ratio, $\Phi_{\rm PDR}$ \citep{Parker17b}. The paper is organised as follows. In Section~2 we briefly describe the data, in Section~3 we describe the methods used to quantify the spatial distributions, in Section~4 we present our results, we provide a discussion in Section 5 and we conclude in Section~6. We also provide an Appendix (A) to discuss different methods of normalising the \citet{Cartwright04} $\mathcal{Q}$-parameter.





\section{Data}

\begin{figure*}
\begin{center}
\rotatebox{270}{\includegraphics[scale=0.75]{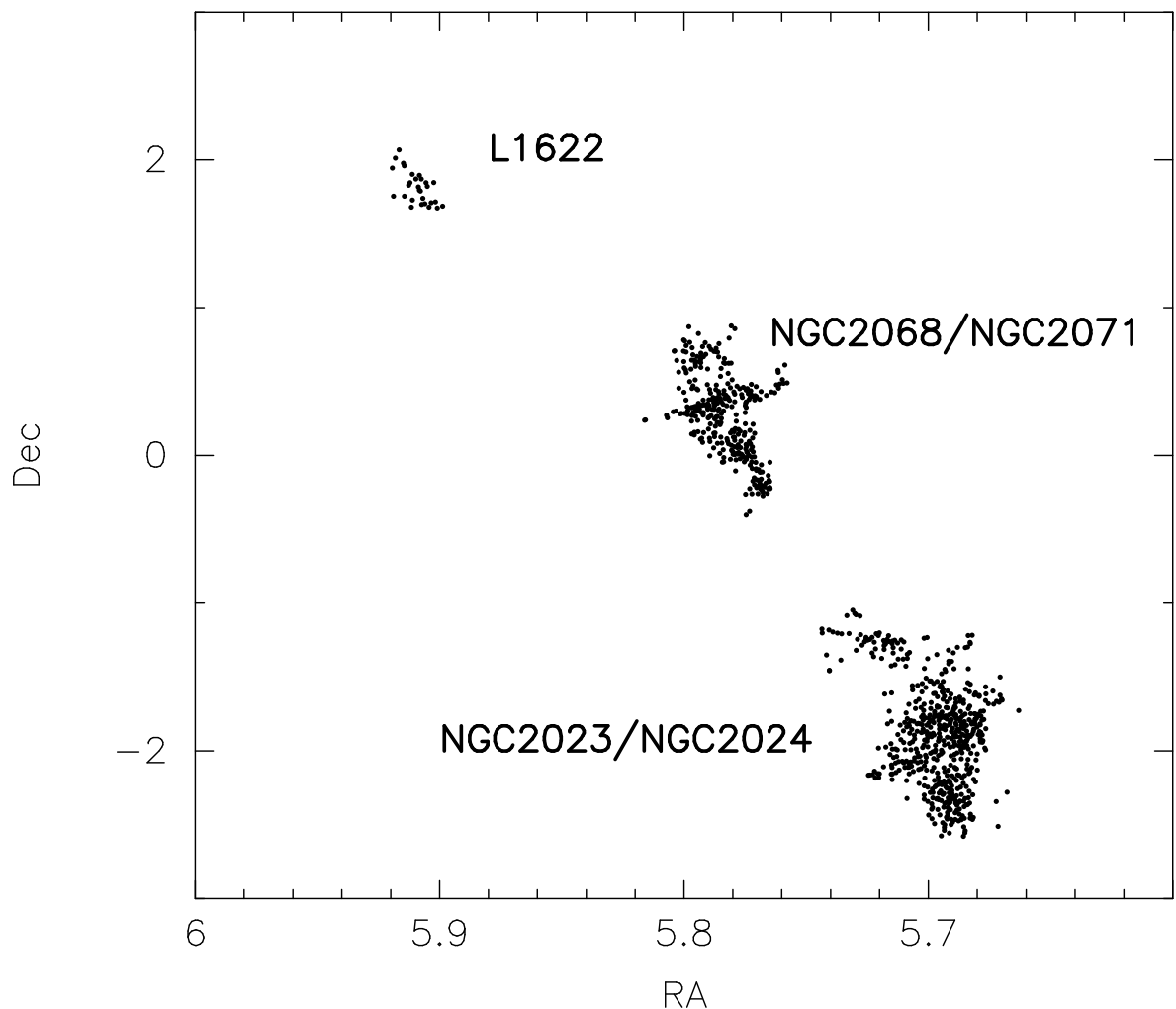}}
\end{center}
\caption[bf]{Map of Orion~B, showing the location of cores in the three spatially distinct regions.}
\label{Full_map}
\end{figure*}

We use the same dataset as \citet{Kirk16b}, namely the James Clerk Maxwell Telescope (JCMT) Gould Belt Survey data on Orion~B, taken with the SCUBA--2 instrument. This dataset comprises a total of 915 prestellar cores split into three spatially distinct star-forming regions, Linds~1622, NGC\,2068/NGC\,2071 and NGC\,2023/NGC\,2024. L1622 contains 29 cores, NGC\,2068/NGC\,2071 contain 322 and NGC\,2023/NGC\,2024 contain 564 cores. We follow \citet{Kirk16b} by adopting the 850$\mu$m flux as a proxy for the masses of the individual cores. The positions of the individual cores are shown in Fig.~\ref{Full_map}.

\section{Methods}

In this section we describe the four diagnostics used to quantify the spatial distribution of dense cores in the data.

\subsection{The $\mathcal{Q}$-parameter}

The $\mathcal{Q}$-parameter was introduced by \citet{Cartwright04} to distinguish between substructured or self-similar (e.g. fractal) distributions, and smooth or centrally concentrated (e.g. clustered) distributions, and has been extensively utilised \citep[e.g.][]{Schmeja06,Bastian09,Gutermuth09,Cartwright09a,Cartwright09b,Sanchez09,Lomax11,Delgado13,Parker12d,Parker14b,Jaffa17,Dib18}. It employs a graph theory approach by constructing a minimum spanning tree (MST), which connects all of the points in a given distribution via the shortest possible path with no closed loops. The mean MST edge length, $\bar{m}$ is determined, and is then normalised by dividing by the following factor which depends on both the number of points $N$, and the area $A$:
\begin{equation}
\frac{\sqrt{NA}}{N - 1}.
\end{equation}
The area, $A$, is taken by \citet{Cartwright04} to be the area of a circle with radius $R$, which encompasses the furthest point from the centre of the distribution. The mean separation length between all of the points in the distribution, $\bar{s}$ is then determined and is normalised to the radius $R$ of the circle.

The normalisation means that $\mathcal{Q}$ is independent of the extent of the region under investigation, and enables a comparison to be made between the spatial properties of different observed and simulated star-forming regions. Several modifications to the original normalisation of $\mathcal{Q}$ have been proposed, and we highlight two here. Firstly, \citet{Schmeja06} replaced the area $A$ with the area of a convex hull $A_{\rm CH}$; a closed set of lines that encompass the outermost points in a distribution. They then normalise $\bar{s}$ to the radius of a circle with the area of this convex hull, $R_{\rm CH-circ}$. Secondly, \citet{Kirk16b} also use the convex hull area $A_{\rm CH}$ to normalise  $\bar{m}$, but then use the distance between the centre of the convex hull and the most distant point from this centre, $R_{\rm CH-ex}$ to normalise $\bar{s}$. In Appendix~\ref{appendix} we compare the three normalisation methods and find the full convex hull method adopted by \citet{Kirk16b} to be flawed for the determination of $\mathcal{Q}$.

Interpreting the calculated value for the $\mathcal{Q}$-parameter requires a comparison with synthetic star-forming regions (i.e.\,\,distributions of points). These are usually either centrally concentrated distributions with a radial density profile of the form $n \propto r^{-\alpha}$, with $\alpha$ in the range 0 -- 3.0, or substructured distributions, with varying levels of substructure described by a fractal distribution with a notional fractal dimension, $D$.

  We construct fractals using the box method described in \citet{Goodwin04a} and \citet{Cartwright04}, where a first-generation parent is placed at the centre of a cube of side $N_{\rm div}$ which then spawns $N_{\rm div}$ subcubes, each with a first-generation child at its centre. The fractal is then built by determining which of the children themselves become parents, and spawn their own offspring. This is determined by the fractal dimension, $D$, where 
the probability that the child becomes a parent is given by $N_{\rm div}^{D-3}$. For a lower fractal dimension fewer children mature and the final
distribution contains more substructure.

We note that the fractal distributions created using the box method are often not perfectly self-similar, and some deviation in the amount of substructure from the desired fractal dimension can occur \citep[and this fractal dimension may also differ from a fractal dimension calculated by an alternative means, such as the perimeter--area method, e.g.][]{Cartwright06}. For this reason, in the following analysis we do not assign a fractal dimension to our calculated $\mathcal{Q}$-parameters, and any such fractal dimension would be purely notional.

  Other, more complex distributions can be used as a comparison, but this can lead to an almost infinite amount of parameter space to consider \citep{Bate98b,Parker12d,Jaffa17}. We therefore restrict our comparison to either box fractals as defined by \citet{Goodwin04a,Cartwright04} or centrally concentrated clusters with different radial density profiles \citep{Cartwright04,Cartwright09b}.

\subsection{The mass segregation ratio, $\Lambda_{\rm MSR}$}

Minimum Spanning Trees (MSTs) are often used to quantify the relative spatial distribution of the most massive stars in a star-forming region  \citep{Allison09a,Parker15b}, but the method can be applied to any distribution of points with assigned masses (or indeed any other scalar property), and we will apply it to the dense cores in Orion B.  For the dataset we use in this paper, the `mass  segregation ratio' ($\Lambda_{\rm MSR}$) is defined as the ratio between the average MST pathlength of 10 randomly chosen cores in a star-forming region and 
and that of the 10 most massive cores:
\begin{equation}
\Lambda_{\rm MSR} = {\frac{\langle l_{\rm average} \rangle}{l_{\rm 10}}} ^{+ {\sigma_{\rm 5/6}}/{l_{\rm 10}}}_{- {\sigma_{\rm 1/6}}/{l_{\rm 10}}}.
\end{equation}
As described in \citet{Allison09a,Parker11b}, we define the lower (upper) uncertainty  as the MST
length which lies 1/6 (5/6) of the way through an ordered list of all
the random lengths (corresponding to a 66 per cent deviation from  the
median value, $\langle l_{\rm average} \rangle$). This determination
prevents a single outlying object from heavily influencing the
uncertainty, which could be an issue if using the Gaussian dispersion as the uncertainty estimator.

If $\Lambda_{\rm MSR} > 1$, then the most massive cores are more spatially concentrated than the average cores, and we designate this as significant if the lower error bar also exceeds unity \citep[see also][]{Alfaro16,Gonzalez17}. \citet{Parker15b} show that  $\Lambda_{\rm MSR}$ can sometimes be too sensitive in that it sometimes finds that random fluctuations in low-number distributions lead to  mass segregation according to our definition. Therefore, if $\Lambda_{\rm MSR}$ is calculated to be less than 2, then we also do not consider this to be a significant deviation from a random distribution.

\subsection{The local surface density ratio, $\Sigma_{\rm LDR}$} 

We calculate the relative local surface density of the most massive cores compared to lower-mass cores using the local surface density ratio, $\Sigma_{\rm LDR}$ \citep{Maschberger11,Kupper11,Parker14b}. We first determine the local surface density around each core, $\Sigma$ as
\begin{equation}
\Sigma = \frac{N - 1}{\pi r_N^2},
\end{equation}
where $r_N$ is the distance to the $N^{\rm th}$ nearest neighbour, $N$ \citep{Casertano85}.  We adopt $N = 10$ throughout this work. 

We divide the median $\Sigma$ for the ten most massive cores, $\tilde{\Sigma}_\mathrm{10}$ by the median value for all the cores $\tilde{\Sigma}_\mathrm{all}$ to define a `local density ratio', $\Sigma_{\rm LDR}$ \citep{Parker14b}:
\begin{equation}
\Sigma_{\rm LDR} = \frac{\tilde{\Sigma}_\mathrm{10}}{\tilde{\Sigma}_\mathrm{all}}.
\end{equation} 
If $\Sigma_{\rm LDR} > 1$ then the most massive cores are in areas of higher local surface density than the average core, the significance of which is quantified by a Kolmogorov-Smirnov (KS) test on the cumulative distribution of the cores, ranked by their local surface densities $\Sigma$.  We reject the hypothesis that the two subsets are drawn from the same underlying distribution if the KS p-value is less than 0.1.

\subsection{The potential difference ratio, $\Phi_{\rm PDR}$}

\citet{Parker17b} use a method analagous to the local surface density ratio to quantify the difference between the gravitational potential of the most massive cores and the average gravitational potential of all cores. We first determine the local gravitational potential, $\Phi_j$, for each core in the simulation:
\begin{equation}
\Phi_j = -\sum{\frac{m_i}{r_{ij}}},
\end{equation}
where $m_i$ is the mass of the $i^{\rm th}$ core in the summation, and $r_{ij}$ is the distance to the $i^{\rm th}$ core. In a similar analysis to the surface density -- mass distribution $\Sigma_{\rm LDR}$ method \citep[][see above]{Maschberger11}, we plot $\Phi_j$ against $m_j$ for each core.

The potential difference ratio, PDR, is defined as:
\begin{equation}
\Phi_{\rm PDR} = \frac{\tilde{\Phi}_{10}}{\tilde{\Phi}_{\rm all}},
\end{equation}
where $\tilde{\Phi}_{10}$ is the median potential of the 10 most massive cores, and $\tilde{\Phi}_{\rm all}$ is the median potential of the entire region in question. If $\Phi_{\rm PDR} > 1$ then the most massive cores sit in deeper local gravitational potentials than the average core and we quantify the significance of this by means of a KS test on the cumulative distribution of the cores, ranked by their potentials, where we reject the hypothesis that the two subsets are drawn from the same underlying distribution if the KS p-value is less than 0.1. 

\begin{figure}
\begin{center}
\rotatebox{270}{\includegraphics[scale=0.4]{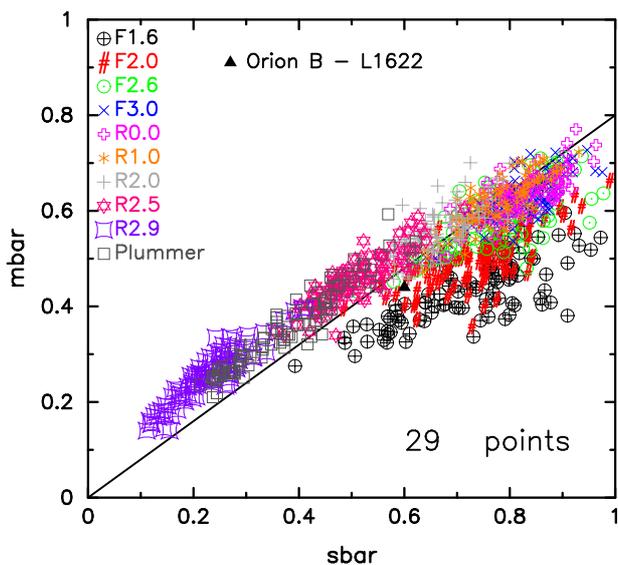}}
\end{center}
\caption[bf]{The location of L1622 on the \citet{Cartwright09b} $\bar{m} - \bar{s}$ plot and compared to synthetic star-forming regions containing the same number of objects (29) as L1622. We show the results for ten different geometries, starting with very substructured fractal regions with fractal dimension $D = 1.6$ (the black $\oplus$ symbols) and increasing in fractal dimension (corresponding to increasingly smoother distributions) until the fractals produce a uniform sphere ($D = 3.0$, the blue crosses). We then switch regimes to regions that are smooth and centrally concentrated with a radial density profile $n \propto r^{-\alpha}$, where $\alpha = 0$ indicates a uniform density profile, up to $\alpha = 2.9$ (the purple squashed squares). We also show the results for Plummer spheres (open charcoal squares). The boundary between substructured and smooth distributions is shown by the solid black line. We show 100 realisations of each geometry.}
\label{L1622_Q}
\end{figure}

\begin{figure*}
  \begin{center}
\setlength{\subfigcapskip}{10pt}
\hspace*{-1.5cm}\subfigure[L1622 map]{\label{L1622-a}\rotatebox{270}{\includegraphics[scale=0.4]{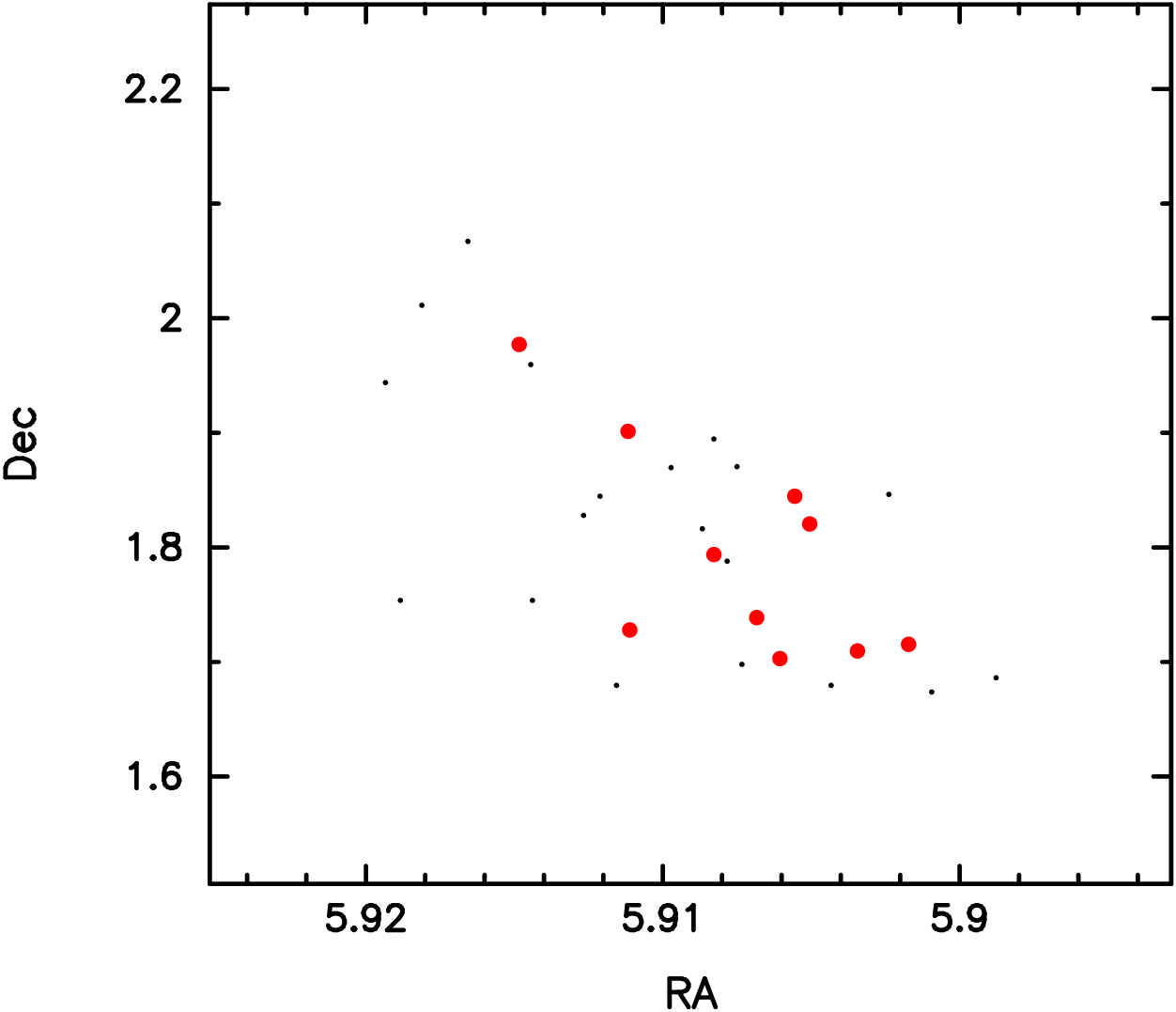}}}
\hspace*{0.3cm} 
\subfigure[$\Lambda_{\rm MSR}$]{\label{L1622-b}\rotatebox{270}{\includegraphics[scale=0.38]{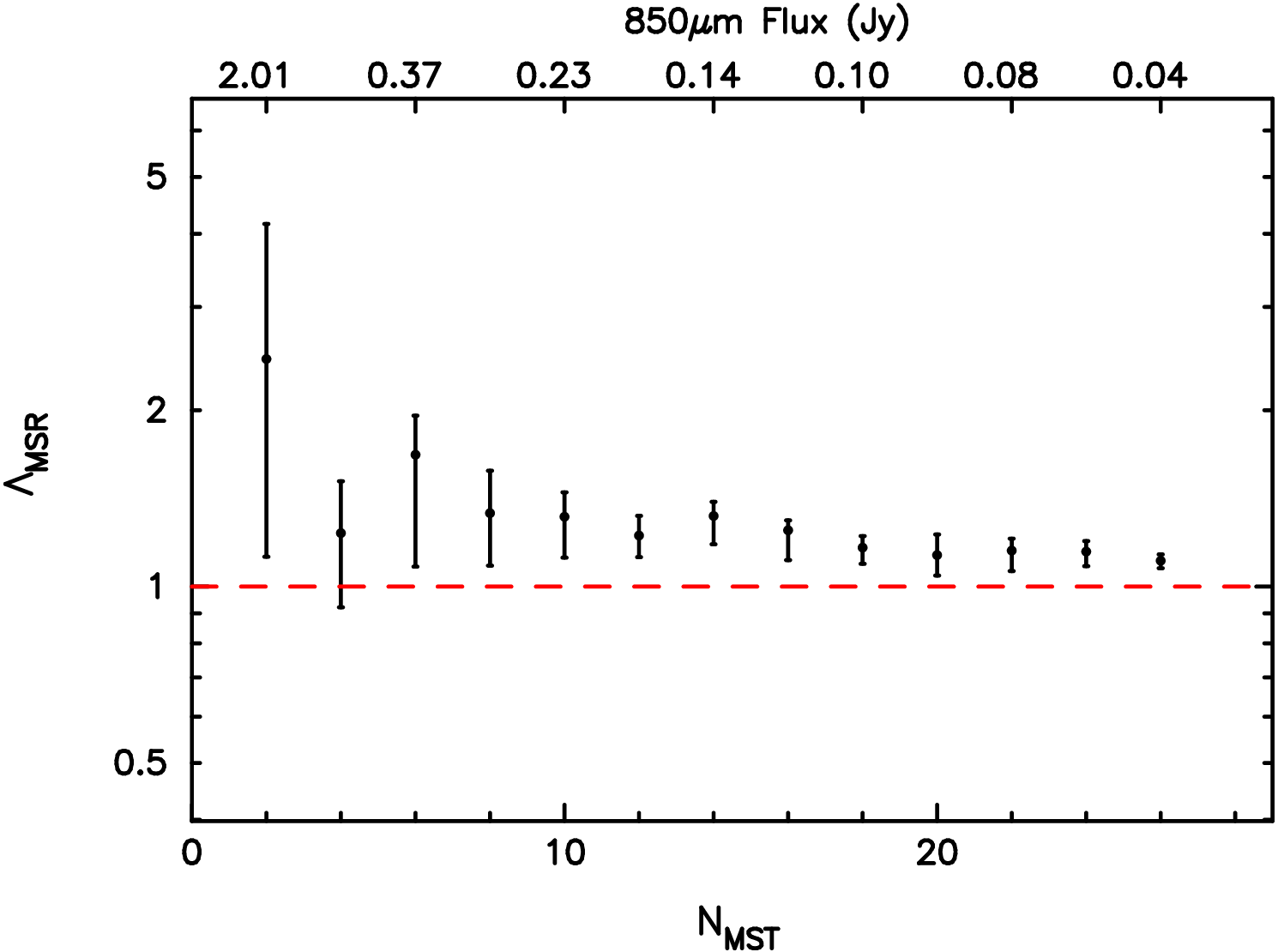}}} 
\hspace*{-1.5cm}\subfigure[$\Sigma_{\rm LDR}$]{\label{L1622-c}\rotatebox{270}{\includegraphics[scale=0.40]{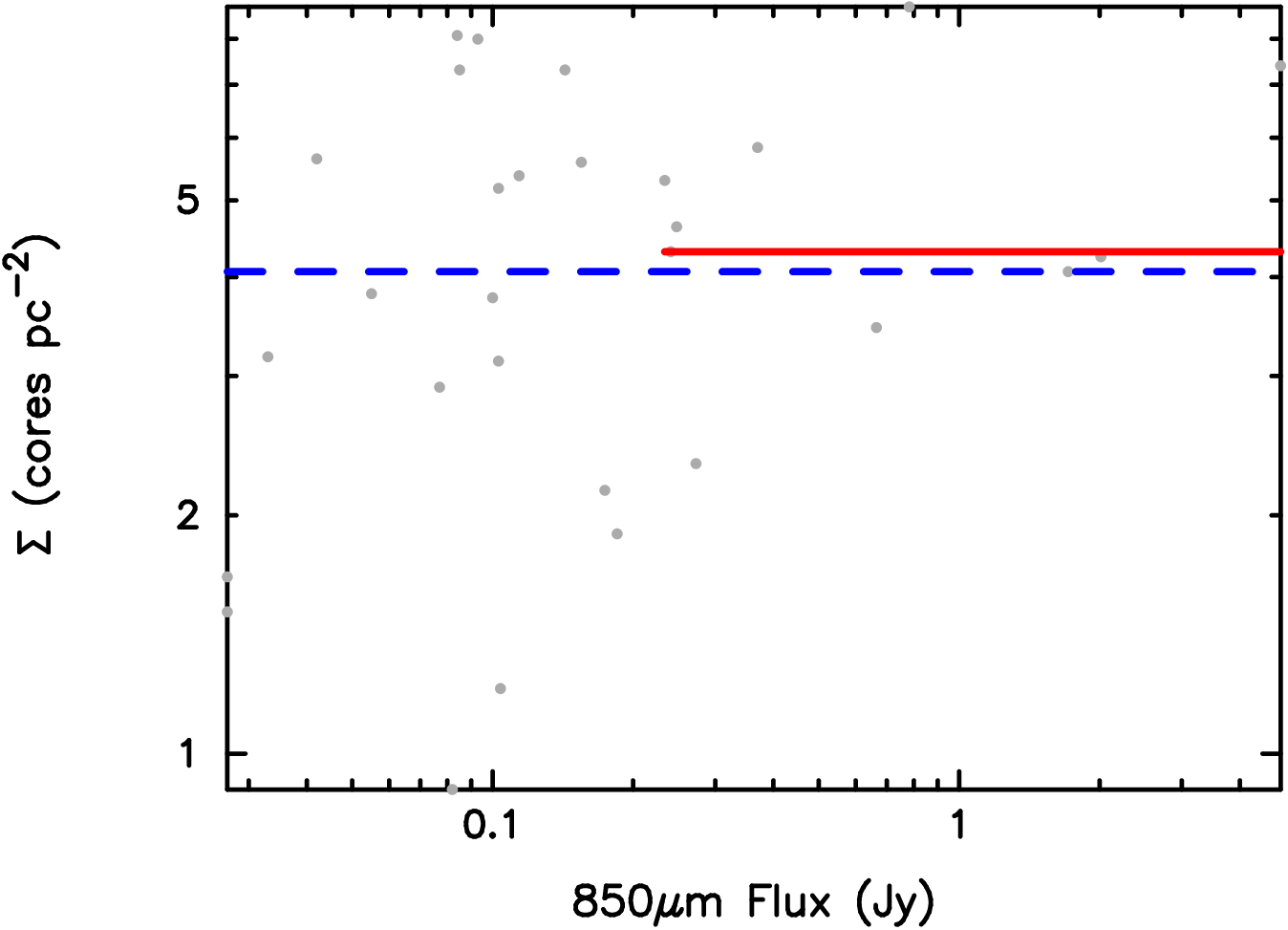}}}
\hspace*{0.3cm} 
\subfigure[$\Phi_{\rm PDR}$]{\label{L1622-d}\rotatebox{270}{\includegraphics[scale=0.40]{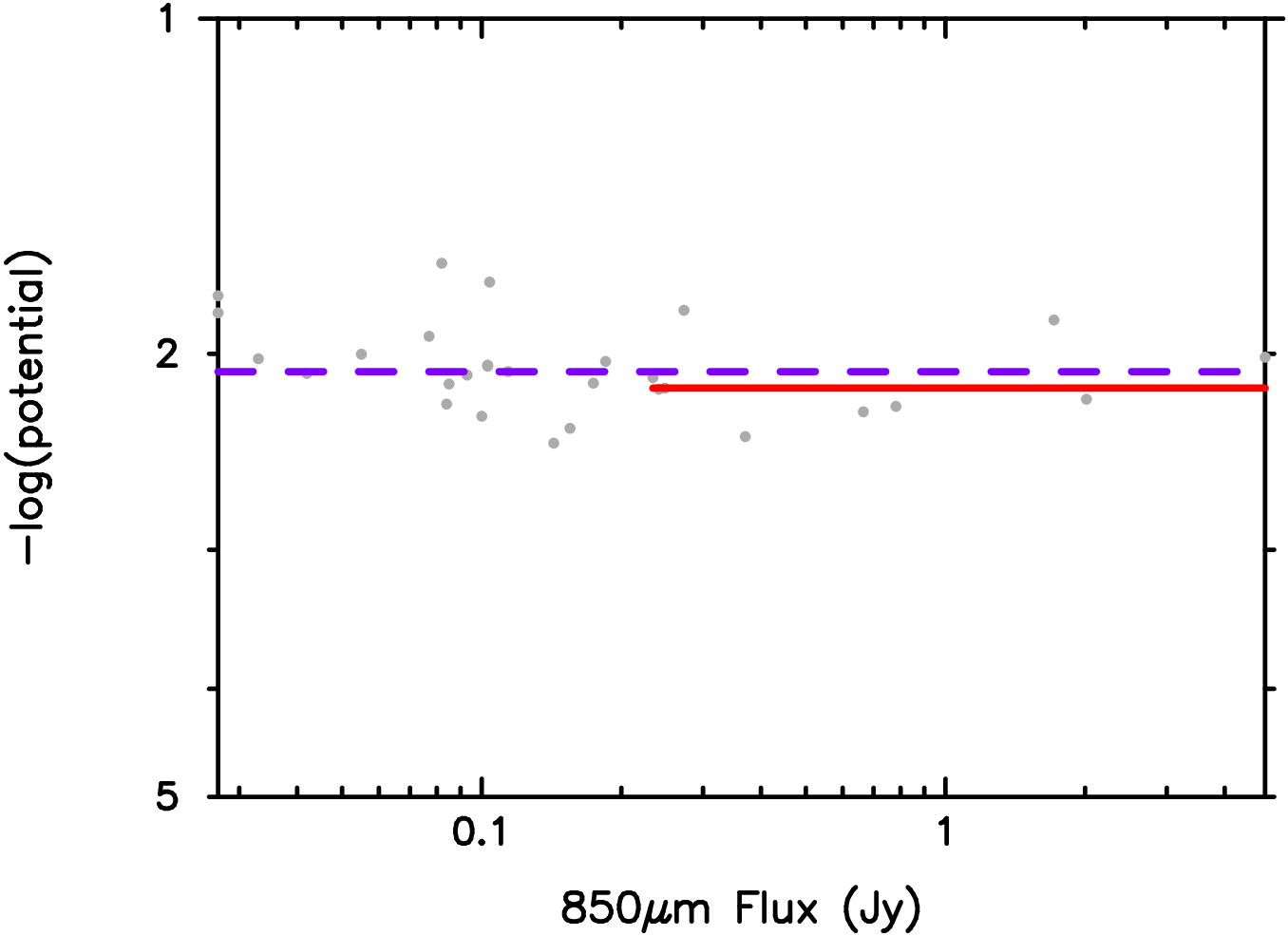}}}
\caption[bf]{Spatial distributions of the most massive cores (i.e.\,\,those with the highest 850$\mu$m flux) in L1622. In panel (a) we show the location of the most massive cores (the red points). In panel (b) we show the mass segregation ratio, $\Lambda_{\rm MSR}$ as a function of the $N_{\rm MST}$ cores, ordered by decreasing 850$\mu$m flux. The dashed line indicates $\Lambda_{\rm MSR} = 1$, corresponding to no mass segregation. In panel (c) we show the local surface density $\Sigma$ as a function of the individual 850$\mu$m flux of each core. The solid red line indicates the median surface density for the ten most massive cores, and the blue dashed line indicates the median $\Sigma$ value for the entire  L1622 region. Finally, in panel (d) we show the local gravitational potential, $\Phi$ as a function of the individual 850$\mu$m flux of each core. The solid red line shows the median $\Phi$ value for the ten most massive cores, and the purple dashed line shows the median  $\Phi$ value for all cores in the region.}
\label{L1622}
  \end{center}
\end{figure*}

\section{Results}
\label{results}

In this section, we follow the approach of \citet{Kirk16b} and split the Orion B region into its three spatially distinct (in two dimensions) regions: L1622, NGC\,2068/NGC\,2071 and NGC\,2023/NGC\,2024. We then apply the $\mathcal{Q}$-parameter, $\Lambda_{\rm MSR}$ ratio, $\Sigma_{\rm LDR}$ technique and the $\Phi_{\rm PDR}$ technique to the three regions.

\subsection{L1622}

Using the original \citet{Cartwright04} method, we determine a $\mathcal{Q}$-parameter of 0.72, which straddles the boundary between a substructured and a smooth distribution.  In  Fig.~\ref{L1622_Q} we show the \citet{Cartwright09b} $\bar{m} - \bar{s}$ plot, which further distinguishes between the substructured and smooth regimes. Whilst L1622 is marginally in the substructured regime, the small number of cores in this region (29) mean that any interpretation based on these values should be treated with caution.

Interestingly, \citet{Kirk16b} obtain a much higher value for the $\mathcal{Q}$-parameter ($\mathcal{Q} = 1.18$), which would definitively place it in the smooth structural regime. However, we believe there is a flaw in their method used to normalise both  $\bar{m}$ and $\bar{s}$ (and therefore $\mathcal{Q}$ itself), which we discuss in the Appendix of this paper.

Next, we examine the relative distribution of the most massive cores (as defined by their 850$\mu$m flux). In Fig.~\ref{L1622-a} we show the positions of the cores in L1622, highlighting the positions of the ten cores with the highest flux in red.

We show the evolution of the $\Lambda_{\rm MSR}$ mass segregation ratio as a function of the $N_{\rm MST}$ most massive cores in Fig.~\ref{L1622-b}. As with the determination of the $\mathcal{Q}$-parameter, the low number of cores in this region precludes the drawing of any strong conclusions, but we note that the most massive cores do not appear to be significantly more concentrated than lower-mass cores in the region.

The local surface density ratio, $\Sigma_{\rm LDR}$, is marginally above unity (compare the solid red and dashed blue lines in Fig.~\ref{L1622-c}), but a KS-test between the local surface density distribution of the ten most massive cores and the full distribution of all 29 cores returns a KS difference of 0.25 and a p-value of 0.68 that they share the same underlying distribution. 

Finally, the median potential of the most massive cores is slightly higher than that of the full region, i.e.\,\, the most massive cores sit in slightly deeper potentials than the average core (Fig.~\ref{L1622-d}). However, the KS-test between the two distributions returns a KS difference of 0.26 and a p-value of 0.65 that they share the same underlying distribution.

\subsection{NGC\,2068/NGC\,2071}

The $\mathcal{Q}$-parameter for the cores in the NGC\,2068 and NGC\,2071 regions is $\mathcal{Q} = 0.65$ \citep[using the original normalisation from][]{Cartwright04}. This indicates a slightly substructured distribution, and is in line with the visual appearance of the region. In contrast, \citet{Kirk16b} report a $\mathcal{Q} = 0.91$, although again, this high value is due to the erroneous convex hull normalisation technique described in the Appendix.

\begin{figure}
\begin{center}
\rotatebox{270}{\includegraphics[scale=0.4]{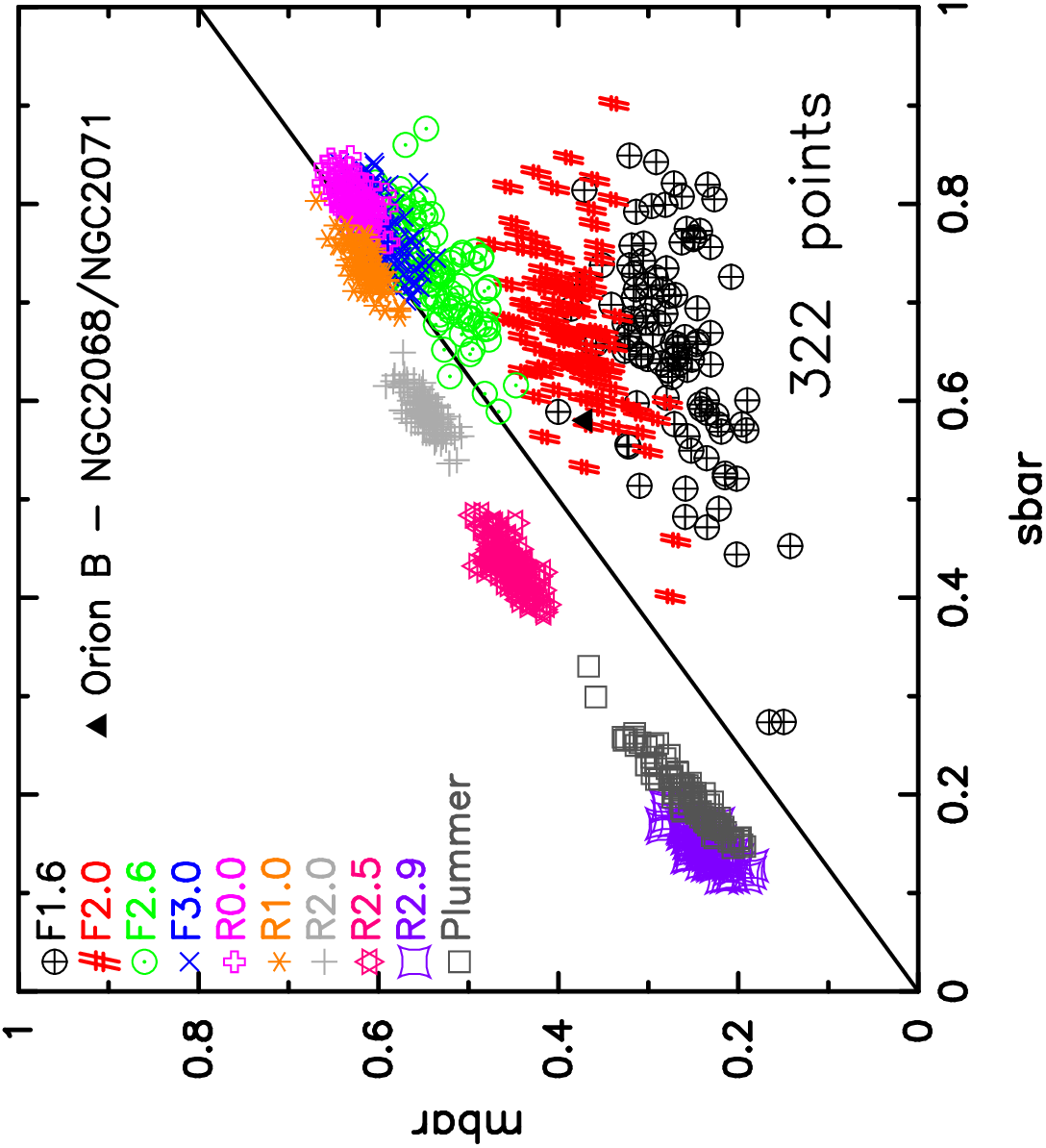}}
\end{center}
\caption[bf]{The location of NGC\,2068/NGC\,2071 on the \citet{Cartwright09b} $\bar{m} - \bar{s}$ plot and compared to synthetic star-forming regions containing the same number of objects (322) as NGC\,2068/2071. We show the results for ten different geometries, starting with very substructured fractal regions with fractal dimension $D = 1.6$ (the black $\oplus$ symbols) and increasing in fractal dimension (corresponding to increasingly smoother distributions) until the fractals produce a uniform sphere ($D = 3.0$, the blue crosses). We then switch regimes to regions that are smooth and centrally concentrated with a radial density profile $n \propto r^{-\alpha}$, where $\alpha = 0$ indicates a uniform density profile, up to $\alpha = 2.9$ (the purple squashed squares). We also show the results for Plummer spheres (open charcoal squares). The boundary between substructured and smooth distributions is shown by the solid black line. We show 100 realisations of each geometry.}
\label{N2068_N2071_Q}
\end{figure}

Unlike L1622, the NGC\,2068/NGC\,2071 region contains enough cores (322) to constrain its spatial distribution using the \citet{Cartwright09b}  $\bar{m} - \bar{s}$ plot. If we place NGC\,2068/NGC\,2071 on the $\bar{m} - \bar{s}$ plot (Fig.~\ref{N2068_N2071_Q}), we see that it resides within the moderately substructured regime and overlaps with the parameter space of fractal distributions with fractal dimension $D = 2.0$. We note that this does not necessarily mean that the distribution of cores in NGC\,2068/NGC\,2071 \emph{is} a fractal distribution, but rather it has the same degree of substructure as a fractal with  $D = 2.0$.

\begin{figure*}
  \begin{center}
\setlength{\subfigcapskip}{10pt}
\hspace*{-1.5cm}\subfigure[NGC\,2068/NGC\,2071 map]{\label{N2068_N2071-a}\rotatebox{270}{\includegraphics[scale=0.4]{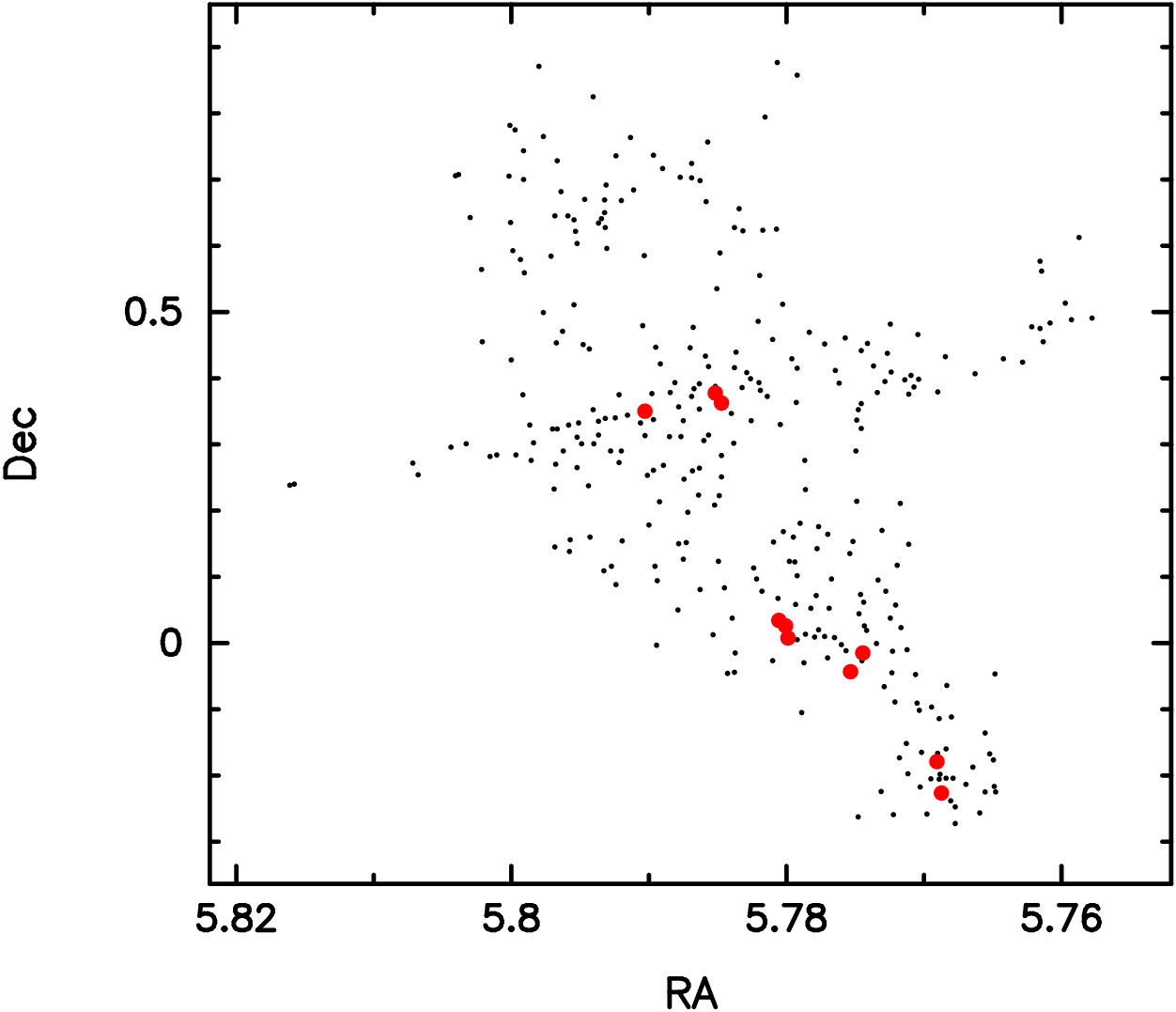}}}
\hspace*{0.3cm} 
\subfigure[$\Lambda_{\rm MSR}$]{\label{N2068_N2071-b}\rotatebox{270}{\includegraphics[scale=0.38]{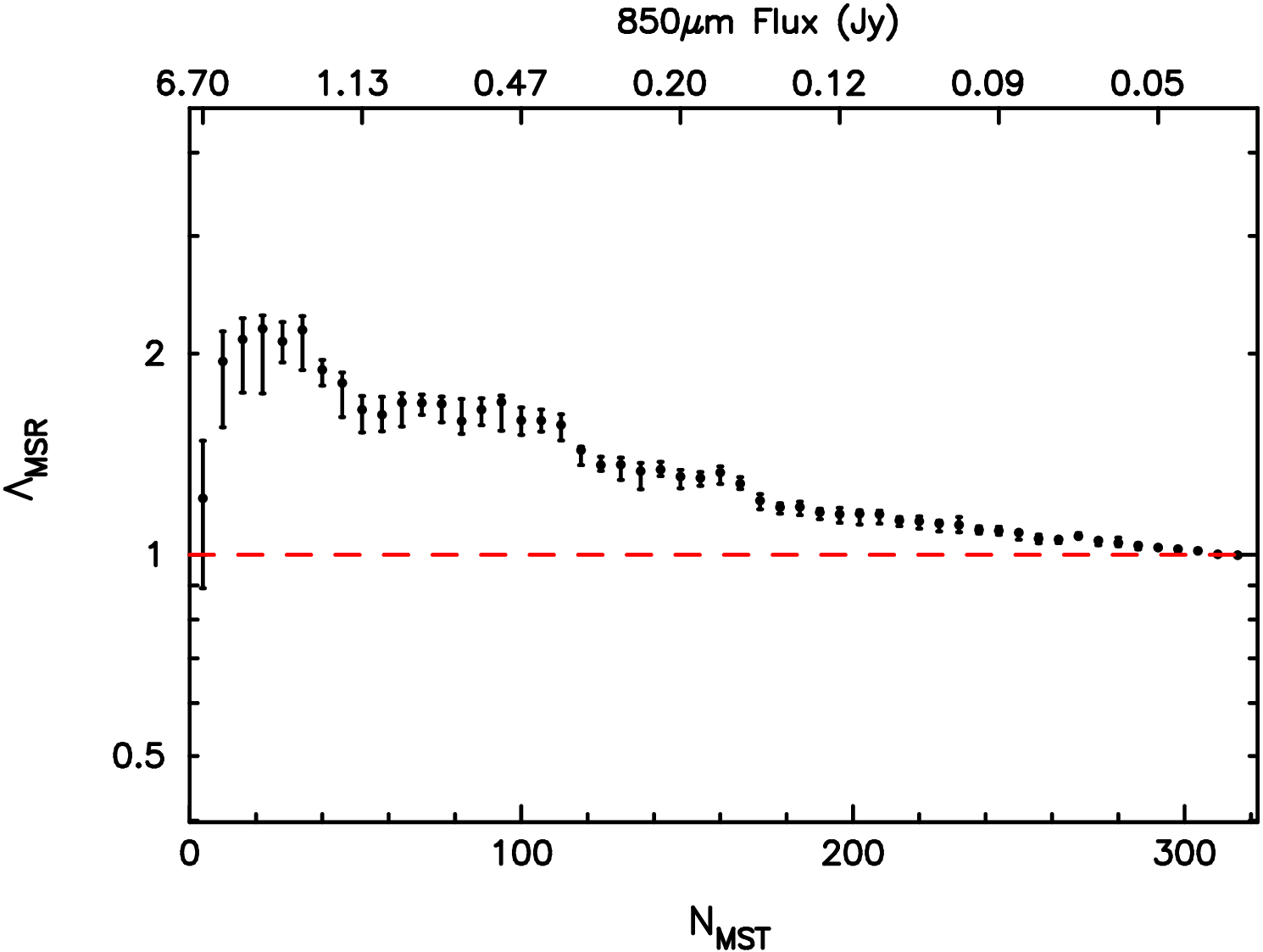}}} 
\hspace*{-1.5cm}\subfigure[$\Sigma_{\rm LDR}$]{\label{N2068_N2071-c}\rotatebox{270}{\includegraphics[scale=0.40]{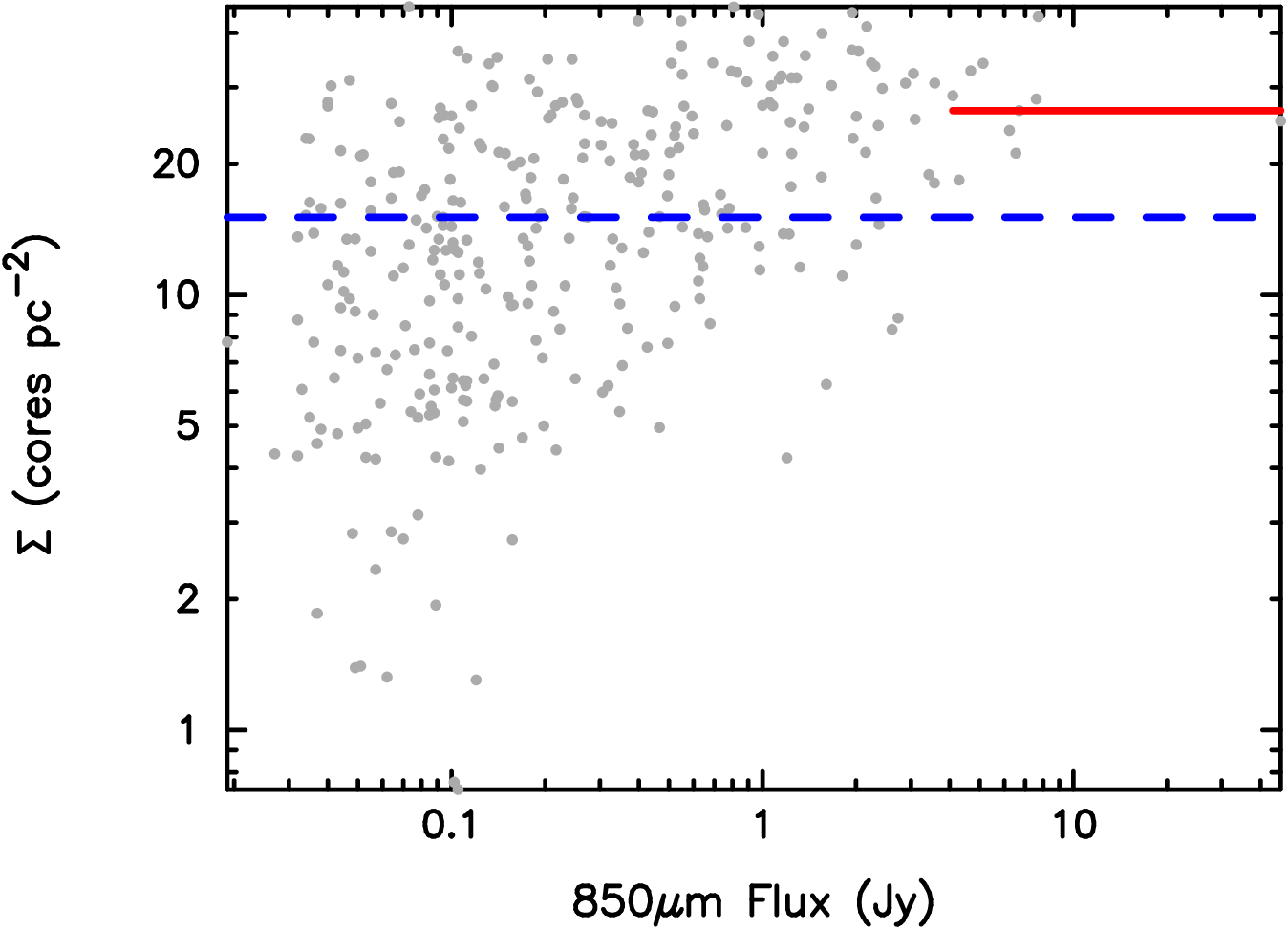}}}
\hspace*{0.3cm} 
\subfigure[$\Phi_{\rm PDR}$]{\label{N2068_N2071-d}\rotatebox{270}{\includegraphics[scale=0.40]{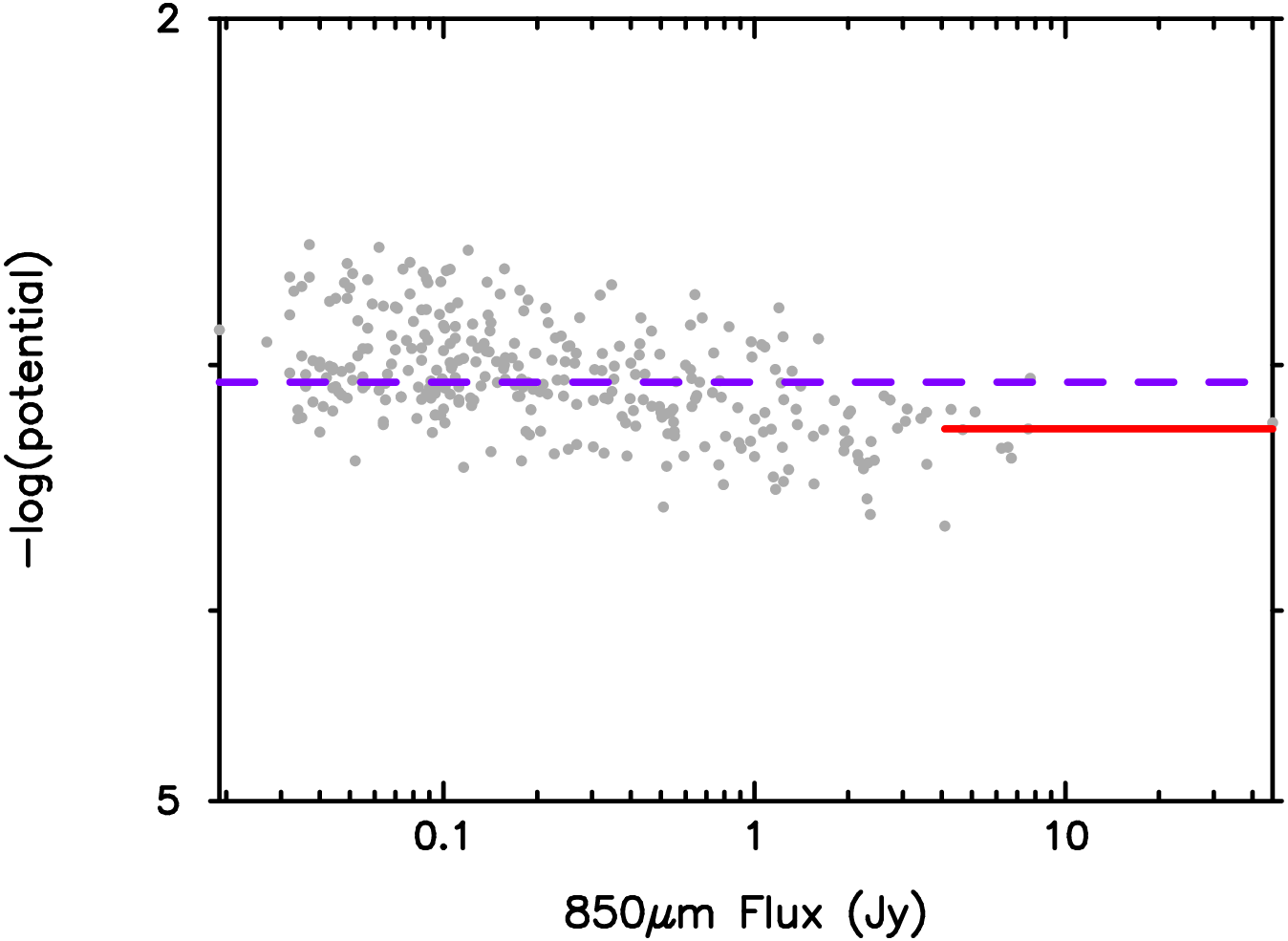}}}
\caption[bf]{Spatial distributions of the most massive cores (i.e.\,\,those with the highest 850$\mu$m flux) in NGC\,2068/NGC\,2071. In panel (a) we show the location of the most massive cores (the red points). In panel (b) we show the mass segregation ratio, $\Lambda_{\rm MSR}$ as a function of the $N_{\rm MST}$ cores, ordered by decreasing 850$\mu$m flux. The dashed line indicates $\Lambda_{\rm MSR} = 1$, corresponding to no mass segregation. In panel (c) we show the local surface density $\Sigma$ as a function of the individual 850$\mu$m flux of each core. The solid red line indicates the median surface density for the ten most massive cores, and the blue dashed line indicates the median $\Sigma$ value for the entire  NCG\,2068/NGC\,2071 region. Finally, in panel (d) we show the local gravitational potential, $\Phi$ as a function of the individual 850$\mu$m flux of each core. The solid red line shows the median $\Phi$ value for the ten most massive cores, and the purple dashed line shows the median  $\Phi$ value for all cores in the region. }
\label{N2068_N2071}
  \end{center}
\end{figure*}

We show the location of the ten most massive cores (those with the highest 850$\mu$m flux) by the large red points in Fig.~\ref{N2068_N2071-a}. The most massive cores appear in groups of two or three, and are distributed over an area which is slightly smaller than the extent of the full region. We quantify the spatial distribution of the most massive cores in Fig.~\ref{N2068_N2071-b} where we show the  $\Lambda_{\rm MSR}$ ratio as a function of the $N_{\rm MST}$ most massive cores. The four most massive cores are consistent with $\Lambda_{\rm MSR} = 1$, whereas the 10 to 40 most massive cores appear significantly more concentrated than the average cores ($\Lambda_{\rm MSR} = 1.95^{+0.2}_{-0.4}$ for the $N_{\rm MST} = 10$ most massive cores).

In Fig.~\ref{N2068_N2071-c} we show the local surface density of the cores in the NGC\,2068/NGC\,2071 region as a function of their 850$\mu$m flux. The median surface density of all cores ($\Sigma = 15$\,cores\,pc$^{-2}$) is shown by the dashed blue line, and the surface density of the ten most massive cores ($\Sigma = 27$\,cores\,pc$^{-2}$) is shown by the solid red line. A KS test between the ten most massive cores and the full region has a KS difference of  0.6 and a p-value of $9 \times 10^{-4}$ that they share the same underlying parent distribution.

The local potential  around each core is shown as a function of 850$\mu$m flux in Fig.~\ref{N2068_N2071-d}. The most massive cores sit in a deeper potential (median $\Phi = -3.23$) than the average cores in the region (median $\Phi = -3.06$). A KS test between the two samples returns a KS difference of 0.62 with a p-value $5.8 \times 10^{-4}$ that they share the same underlying parent distribution.

\subsection{NGC\,2023/NGC\,2024}

Finally, we examine the distribution of 564 cores in the NGC\,2023 and NGC\,2024 regions. The $\mathcal{Q}$-parameter for the cores in these regions is $\mathcal{Q} = 0.71$, which is close to the boundary between a substructured and a smooth distribution. As before, our calculated $\mathcal{Q}$-parameter is lower than that determined by \citet{Kirk16b} using the flawed convex hull normalisation described in the Appendix (they find $\mathcal{Q} = 0.99$).

\begin{figure}
\begin{center}
\rotatebox{270}{\includegraphics[scale=0.4]{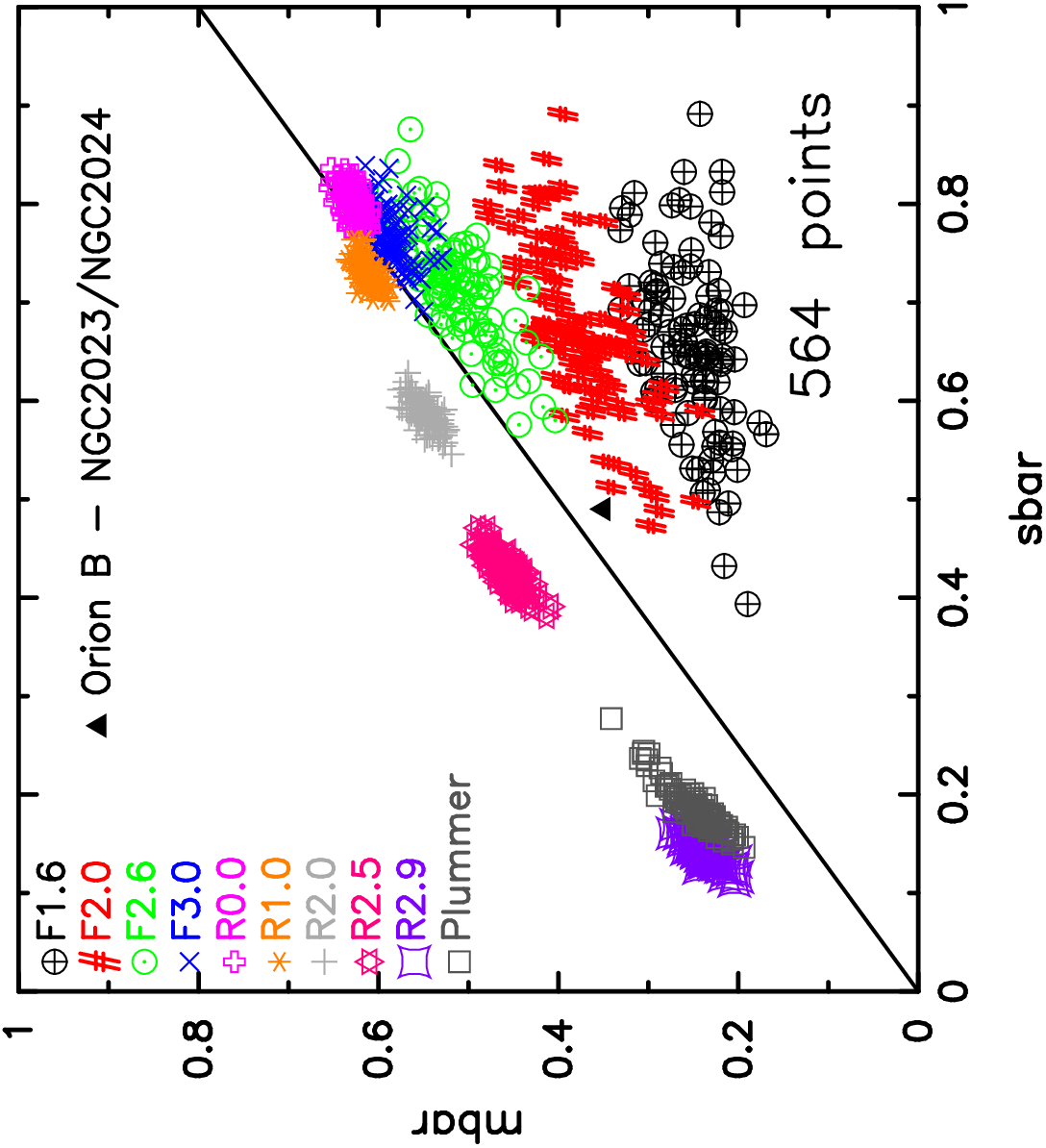}}
\end{center}
\caption[bf]{The location of NGC\,2023/NGC\,2024 on the \citet{Cartwright09b} $\bar{m} - \bar{s}$ plot and compared to synthetic star-forming regions containing the same number of objects (564) as NGC\,2023/2024. We show the results for ten different geometries, starting with very substructured fractal regions with fractal dimension $D = 1.6$ (the black $\oplus$ symbols) and increasing in fractal dimension (corresponding to increasingly smoother distributions) until the fractals produce a uniform sphere ($D = 3.0$, the blue crosses). We then switch regimes to regions that are smooth and centrally concentrated with a radial density profile $n \propto r^{-\alpha}$, where $\alpha = 0$ indicates a uniform density profile, up to $\alpha = 2.9$ (the purple squashed squares). We also show the results for Plummer spheres (open charcoal squares). The boundary between substructured and smooth distributions is shown by the solid black line. We show 100 realisations of each geometry.}
\label{N2023_N2024_Q}
\end{figure}

The $\mathcal{Q}$-parameter calculated using the \citet{Cartwright04} method cannot be used in isolation to determine the structural properties of the NGC\,2023 and NGC\,2024 regions. We show the \citet{Cartwright09b}  $\bar{m} - \bar{s}$ plot in Fig.~\ref{N2023_N2024_Q} for synthetic regions containing 564 points with a range of different morphologies.  NGC\,2023/NGC\,2024 has a similar spatial distribution to a fractal region with $D = 2.0$, but we again emphasise that this does not mean that NGC\,2023/NGC\,2024 \emph{is} a fractal.

\begin{figure*}
  \begin{center}
\setlength{\subfigcapskip}{10pt}
\hspace*{-1.5cm}\subfigure[NGC\,2023/NGC\,2024 map]{\label{N2023_N2024-a}\rotatebox{270}{\includegraphics[scale=0.4]{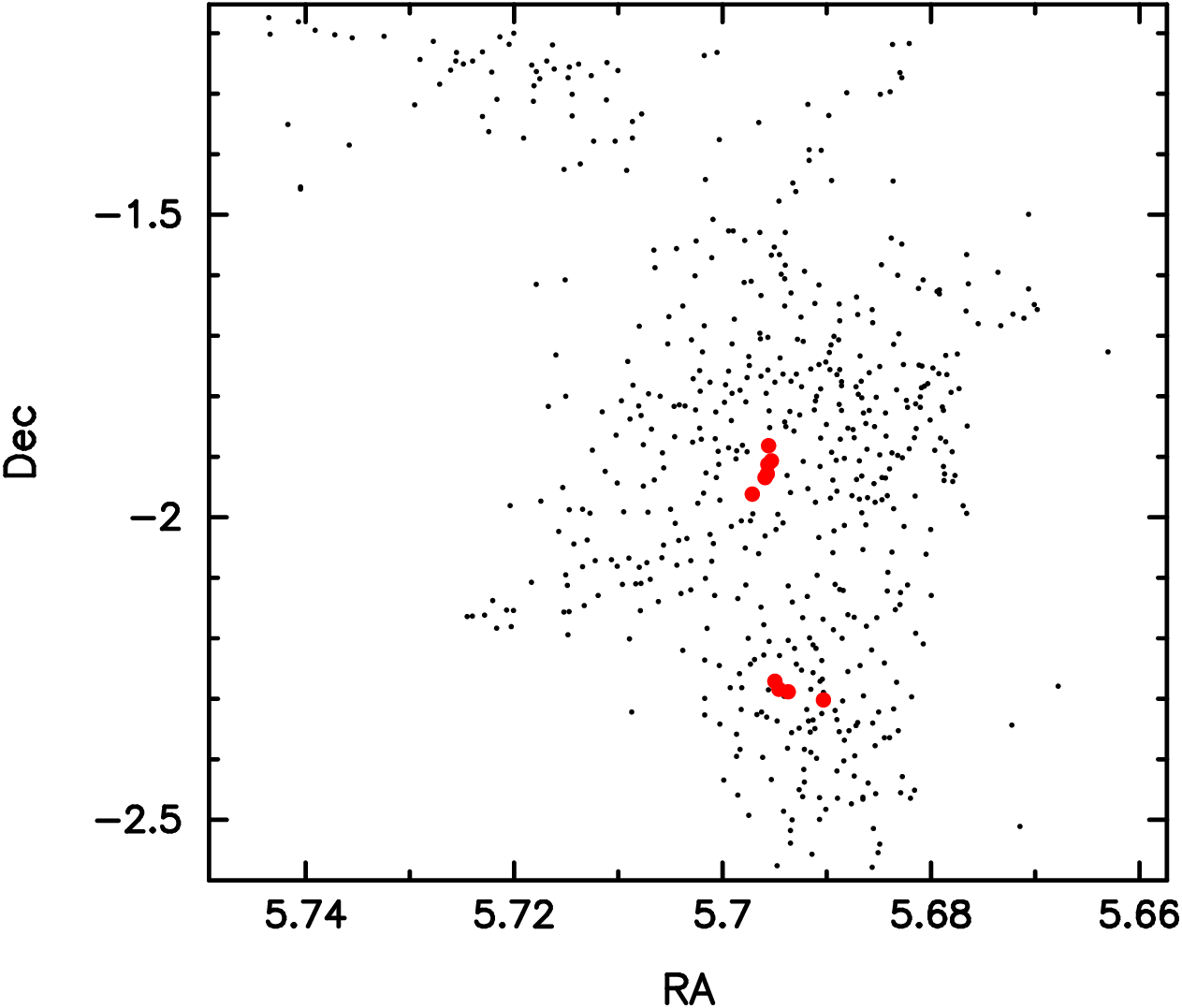}}}
\hspace*{0.3cm} 
\subfigure[$\Lambda_{\rm MSR}$]{\label{N2023_N2024-b}\rotatebox{270}{\includegraphics[scale=0.38]{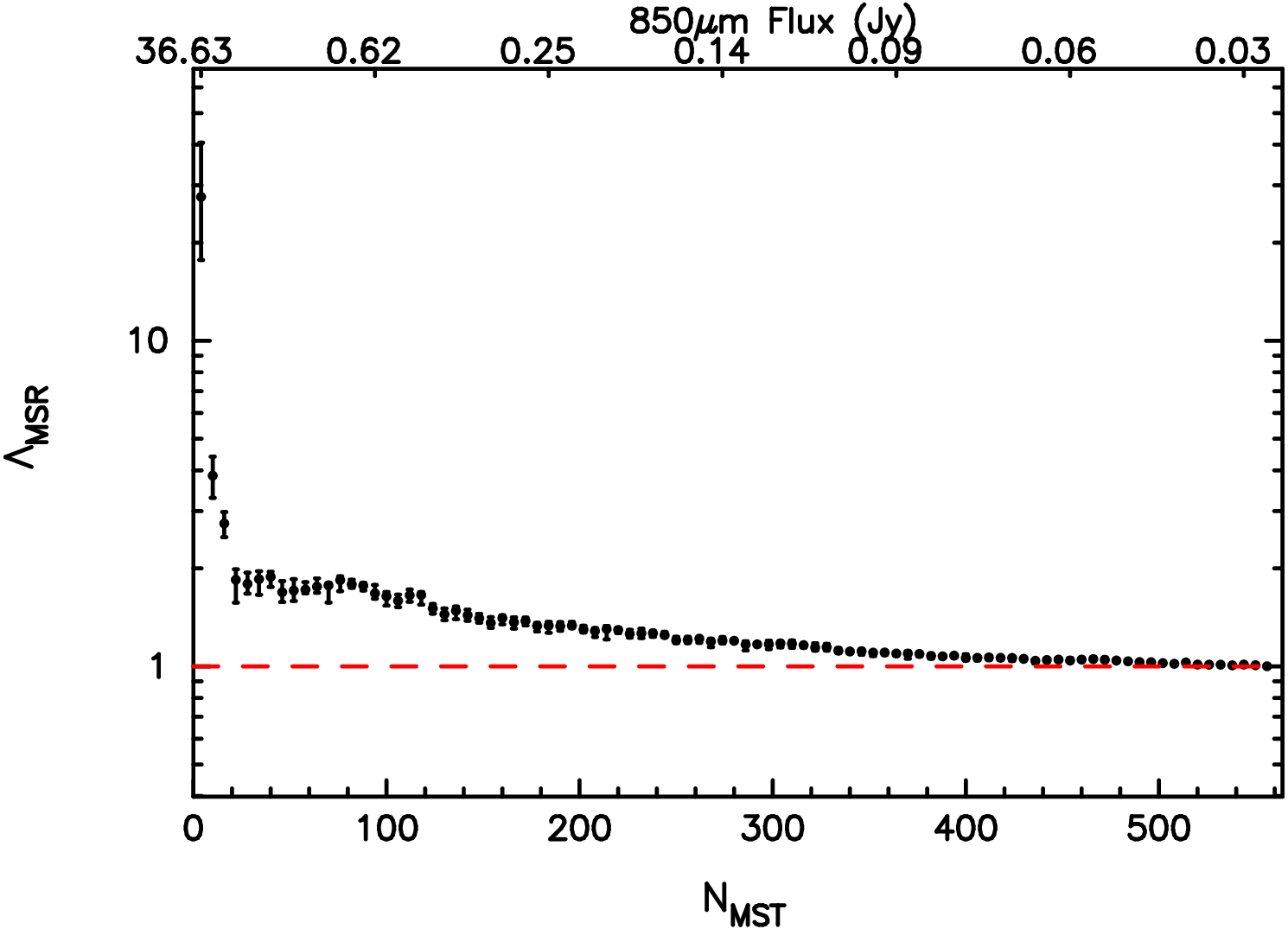}}} 
\hspace*{-1.5cm}\subfigure[$\Sigma_{\rm LDR}$]{\label{N2023_N2024-c}\rotatebox{270}{\includegraphics[scale=0.40]{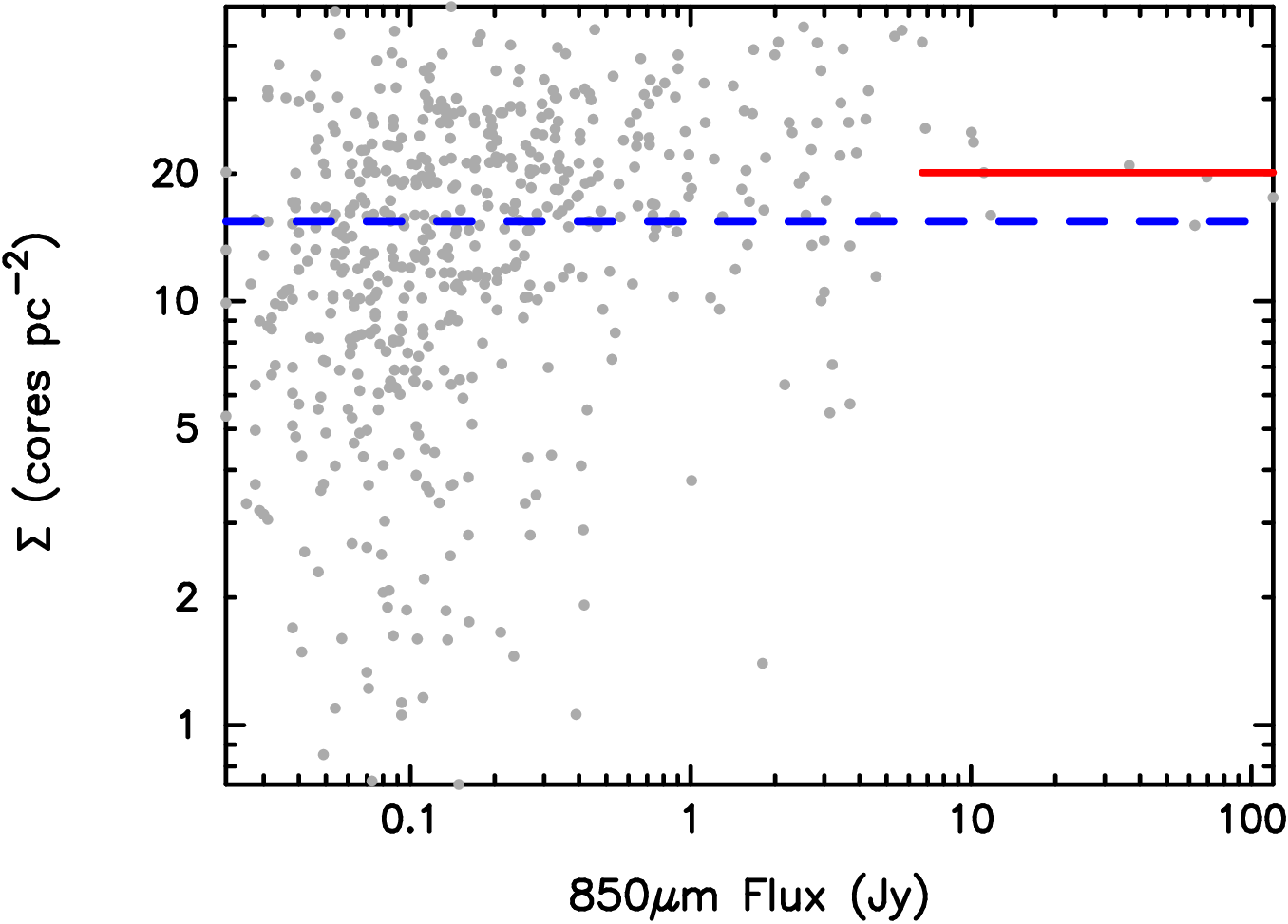}}}
\hspace*{0.3cm} 
\subfigure[$\Phi_{\rm PDR}$]{\label{N2023_N2024-d}\rotatebox{270}{\includegraphics[scale=0.40]{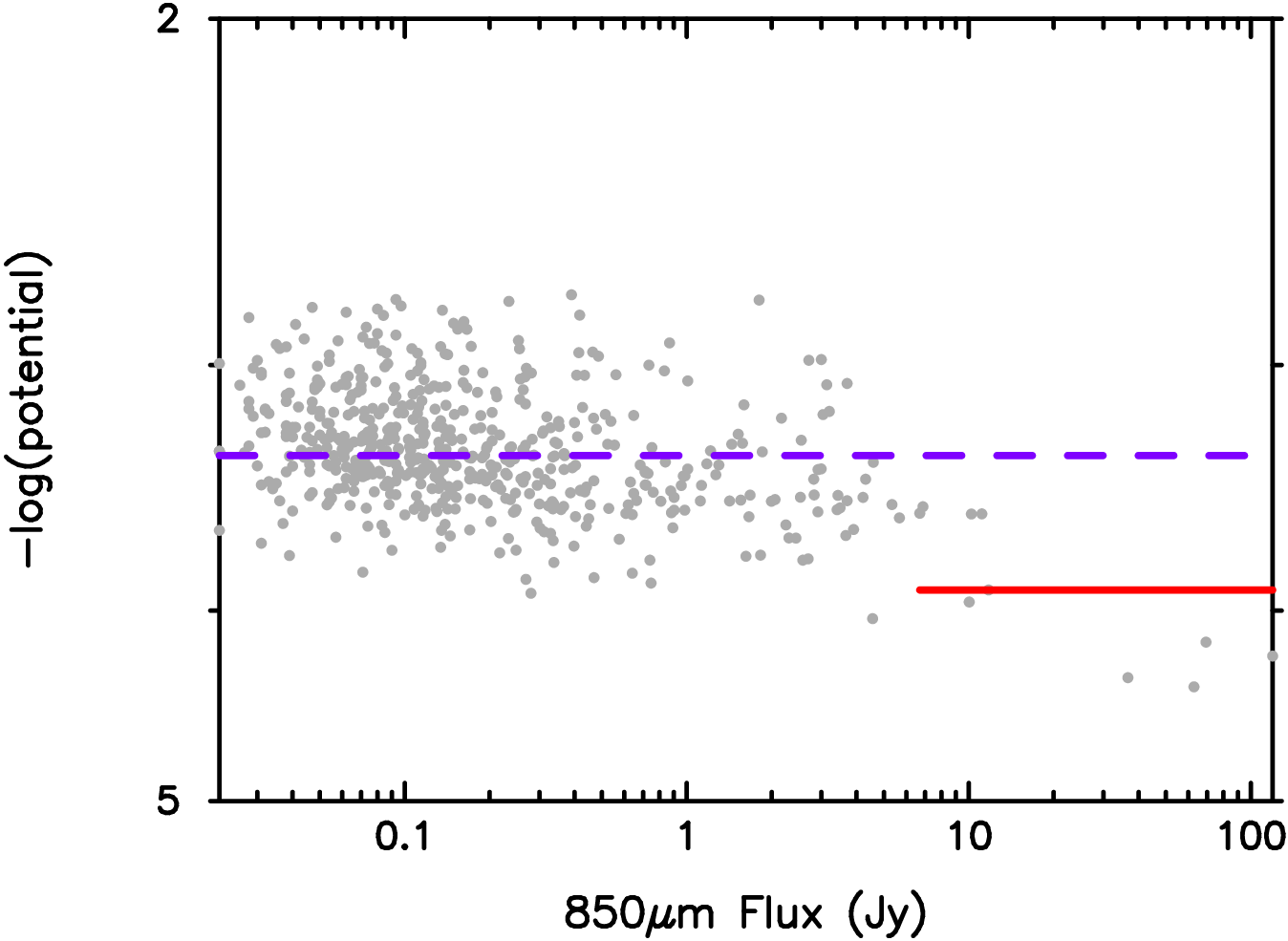}}}
\caption[bf]{Spatial distributions of the most massive cores (i.e.\,\,those with the highest 850$\mu$m flux) in NGC\,2023/NGC\,2024. In panel (a) we show the location of the most massive cores (the red points). In panel (b) we show the mass segregation ratio, $\Lambda_{\rm MSR}$ as a function of the $N_{\rm MST}$ cores, ordered by decreasing 850$\mu$m flux. The dashed line indicates $\Lambda_{\rm MSR} = 1$, corresponding to no mass segregation. In panel (c) we show the local surface density $\Sigma$ as a function of the individual 850$\mu$m flux of each core. The solid red line indicates the median surface density for the ten most massive cores, and the blue dashed line indicates the median $\Sigma$ value for the entire  NCG\,2023/NGC\,2024 region. Finally, in panel (d) we show the local gravitational potential, $\Phi$ as a function of the individual 850$\mu$m flux of each core. The solid red line shows the median $\Phi$ value for the ten most massive cores, and the purple dashed line shows the median  $\Phi$ value for all cores in the region.}
\label{N2023_N2024}
  \end{center}
\end{figure*}

We show the locations of the ten most massive cores (as defined by their  850$\mu$m flux)  by the large red points in Fig.~\ref{N2023_N2024-a}. It is clear that the most massive cores are more clustered than the average cores, and we quantify this using the $\Lambda_{\rm MSR}$ ratio as a function of the $N_{\rm MST}$ most massive cores in Fig.~\ref{N2023_N2024-b}. In contrast to L1622 and NGC\,2068/NGC\,2071, the cores in this region are significantly segregated, with $\Lambda_{\rm MSR} = 28^{+13}_{-10}$ for the $N_{\rm MST} = 4$ most massive cores. The ten most massive cores also display significant mass segregation, with  $\Lambda_{\rm MSR} = 3.9^{+0.5}_{-0.6}$.

Interestingly, the median local surface density of the most massive cores -- whilst significantly higher than the median surface density for all cores -- is not as extreme as the mass segregation measured by $\Lambda_{\rm MSR}$ when compared to NGC\,2068/NGC\,2071. In Fig.~\ref{N2023_N2024-c} we show the local surface density for each core as a function of its 850$\mu$m flux. The median value for the full region ($\Sigma = 15$\,cores\,pc$^{-2}$) is shown by the blue dashed line and the median value for the most massive cores ($\Sigma = 20$\,cores\,pc$^{-2}$) is shown by the solid red line. A KS test on the two samples returns a KS difference of 0.49 and a p-value of $1 \times 10^{-2}$ that they share the same underlying parent distribution.

The local potential  around each core in the NGC\,2023/NGC\,2024 region is shown as a function of 850$\mu$m flux in Fig.~\ref{N2023_N2024-d}. The most massive cores sit in a deeper potential (median $\Phi = -3.9$) than the average cores in the region (median $\Phi = -3.3$). A KS test between the two samples returns a KS difference of 0.85 with a p-value $2.9 \times 10^{-7}$ that they share the same underlying parent distribution.  
 
\section{Discussion}

To summarise our results, we find moderate to low $\mathcal{Q}$-parameters ($\mathcal{Q} < 0.8$) for all three star-forming regions within Orion~B, indicating that these regions are mildly substructured. In L1622, which hosts only 29 cores, the spatial distributions of the most massive cores (as defined by their 850$\mu$m flux) are indistinguishable from the spatial distributions of all cores. However, in NGC\,2068/NGC\,2071 and NGC\,2023/NGC\,2024, the most massive cores reside in areas of higher than average surface density, and sit in deeper potentials than the average core. Interestingly, NGC\,2023/NGC\,2024 displays very high levels of mass segregation from the four most massive cores to the the twenty most massive cores, according to $\Lambda_{\rm MSR}$. The four most massive cores are not mass segregated in the NGC\,2068/NGC\,2071 region, but the 10 -- 40 most massive cores are slightly mass segregated.

\subsection{Caveats and assumptions}

Before discussing these results in the context of star formation theories, and the spatial distributions of pre-main sequence stars in star-forming regions, it is worth highlighting several caveats. First, a single core is unlikely to produce a single star, but rather several during subsequent fragmentation process(es) \citep{Goodwin07,Hatchell08,Lomax14}. It is unclear whether the stars produced by a core would necessarily follow the same spatial distribution as the cores, even if \citep[as proposed by e.g.][]{Alves07} the initial mass function of stars is a direct mapping of the core mass function but at a reduced efficiency.

Secondly, we have followed the procedure of \citet{Kirk16b} and ranked the core masses in terms of their 850$\mu$m flux. If the relation between flux and core mass is not linear, or breaks down in certain regimes, then our determination of $\Lambda_{\rm MSR}$, $\Sigma_{\rm LDR}$ and $\Phi_{\rm PDR}$ could change.

Thirdly, we note that all of the techniques we employ to quantify the spatial distribution of cores ($\mathcal{Q}$, $\Lambda_{\rm MSR}$, $\Sigma_{\rm LDR}$ and $\Phi_{\rm PDR}$) suffer from the same potential biases as when they are applied to quantify the distributions of stars in star-forming regions. For example, if the sample is contaminated by fore- and/or background objects, the $\mathcal{Q}$ parameter will suggest a more homogeneous distribution \citep{Parker12d}, with values tending to $\mathcal{Q} \sim 0.8$. This bias could also have the effect of making the brightest or most massive objects appear more spatially substructured.

Similarly, crowding and extinction in the central regions of star-forming regions can obscure low-mass/low-flux objects, causing the more massive objects to appear more centrally concentrated \citep{Ascenso09,Parker15b}.  However, in such a scenario we would expect the surface density ratio, $\Sigma_{\rm LDR}$ to be lowered, as the massive objects would appear to be relatively isolated if lower-mass objects were obscured.

We note that identifying spatially distinct cores can be difficult in crowded star-forming regions \citep{Kainulainen09}, where choices have to be made on setting the physical boundary of individual cores. This does not affect our comparison with the results of \citet[][see Section 5.2 below]{Kirk16b} because we are using the exact same data, but could affect our determination  of all four of the spatial diagnostics presented in Section 4 and our interpretation of these distributions, which we discuss in Section 5.3.

Finally, we reiterate our point in Section 3.1 that the box fractal method we use to give our calculated $\mathcal{Q}$-parameters physical meaning does not always fully describe the detailed level of substructure in a star-forming region \citep{Jaffa17}. Furthermore, a box fractal with notional fractal dimension $D = 1.6$ will have a much higher (local) density than a fractal with $D = 3.0$ \citep{Bate98b,Parker11c} for the same number of points \citep[see also][]{Lomax11,Parker15c}. However, given the similar dynamic range in both the number of cores \emph{and} local density in Orion B, we do not believe this will negatively impact our interpretation of our calculated $\mathcal{Q}$ values.

\subsection{Comparison with previous work}

Very few studies have quantified the spatial distributions of prestellar cores in star-forming regions. The study by \citet{Kirk16b} was the first to utilise such a large sample of cores and in our study we have used the same dataset as \citet{Kirk16b}, with the same proxy for core mass (850$\mu$m flux). However, due to differences in our adopted methods, our results and intepretation differ significantly.

We find the same behaviour in the surface density--850$\mu$m flux parameter space. All three regions have a low overall density of cores, and the cores with the highest flux tend to be in areas of higher than average surface density.

Our calculated values for the $\mathcal{Q}$-parameter ($\mathcal{Q} = 0.72$ for L1622, $\mathcal{Q} = 0.65$ for NGC\,2068/NGC\,2071 and $\mathcal{Q} = 0.71$ for NGC\,2023/NGC\,2024) differ significantly from those in \citet{Kirk16b} (who report $\mathcal{Q} = 1.18$ for L1622, $\mathcal{Q} = 0.91$ for NGC\,2068/NGC\,2071 and $\mathcal{Q} = 0.99$ for NGC\,2023/NGC\,2024), due to the different normalisation methods. As discussed in Appendix A, we believe the full convex hull normalisation method adopted by \citet{Kirk16b} to be flawed, and we advise against using it in future studies. Whereas the  $\mathcal{Q}$-parameters determined by \citet{Kirk16b} suggest smooth distributions for all three sub-regions of Orion~B, our analysis indicates that they are all spatially substructured.

Using $\Lambda_{\rm MSR}$ we find that L1622 does not exhibit mass segregation of the cores at any significant level. NGC\,2068/NGC\,2071 display some moderate mass segregation for the 10 -- 40 most massive cores (but the 4 most massive cores are not mass segregated). In contrast, NGC\,2023/NGC\,2024 display high levels of mass segregation for the four most massive cores, with the 10 -- 20 most massive cores also mass segregated to a high level.

\citet{Kirk16b} find that all three regions in Orion~B are mass segregated, according to the group segregation method developed by \citet{Kirk10,Kirk14}. This method is very different to conventional methods of defining mass segregation, such as quantifying the change in the IMF as a function of distance from the centre of a star-forming region. Instead of considering the whole star-forming region, the group segregation method divides the region into groups based on a threshold length between objects, This threshold length is determined by drawing an MST of the entire region and then finding a break in the distribution of the branch lengths of the minimum spanning tree. The method then determines whether the most massive object in each group is closer to the centre of the group than the average object, and the group is defined as being mass segregated if this is the case.

\citet{Parker15b} discuss several issues with the group segregation method, two of which we briefly reiterate here. First, the definition of a `group' in this method requires there to be at least ten objects within the threshold MST length of each other. The most massive objects in a region may not even be included in the determination of mass segregation if they are in a relatively isolated location. Secondly, by its very construction, the group segregation method makes a distinction between grouped and ungrouped star formation. However, something that is hierachically substructured (like a young star-forming region) has a continuous distribution over all spatial scales and cannot therefore be split into individual subgroups.

For these reasons, we cannot make a direct comparison between these two methods for finding mass segregation in Orion~B, but note that the $\Lambda_{\rm MSR}$ method measures mass segregation in the more conventional sense (an over-concentration of the most massive objects), whereas the group segregation method has major flaws.

\subsection{Primordial mass segregation?}

If the spatial distribution of the stars that form from the cores in Orion~B follow a similar distribution to the cores, then we would expect that the stars in NGC\,2023/NGC\,2024 (and to a lesser extent NGC\,2068/NGC\,2071) to be mass segregated at very early ages. Given the low surface density of cores and the presence of substructure (as measured by the $\mathcal{Q}$-parameter), it is highly unlikely that the cores have dynamically mass segregated on such short timescales \citep{Parker14b,Dominguez17}. Instead, the observed mass segregation of cores -- subject to the caveats listed above -- is almost certainly primordial, i.e.\,\,the outcome of the star formation process.

The competitive accretion model of star formation \citep{Zinnecker82,Bonnell98b,Bonnell01,Bonnell08} posits that the most massive stars form from Jeans-mass seed objects that accrete more gas than their siblings due to their preferential location in gas-rich areas of the star-forming region. Initially, this theory predicted that the most massive stars should be preferentially centrally concentrated, as they are likely to form in deep potential wells with a large gas reservoir. However, recent analyses of simulations in which massive stars do form from competitive accretion show that this process can occur without the massive stars becoming mass segregated, or residing in areas of higher than average surface density \citep{Parker15a,Parker17b}.

\citet{Parker17b} find that massive stars are preferentially located in deeper potential wells than average stars only if the effects of feedback from the massive stars are switched off in the simulation. When photoionising feedback is switched on, the massive stars do not assume a different spatial distribution to lower-mass stars as they form.

\citet{Pety17} point out that NGC\,2023/NGC\,2024 are in the immediate vicinity of several OB stars surrounded by H{\small II} regions, indicating photoionisation is taking place. Indeed, \citet{Pety17} estimate the mean far-ultra violet (FUV) flux in this region to be 45\,$G_0$, where $G_0 = 1.6 \times 10^{-3}$\,erg\,s$^{-1}$\,cm$^{-2}$ is the typical FUV flux in the interstellar medium \citep{Habing68}. Given this relatively strong FUV radiation field, it is unlikely that the most massive cores have been unaffected by this feedback. We therefore argue that the mass segregation of cores in this region has occurred independently of any competitive accretion process during the formation of stars.

The role of magnetic fields in the star formation process, and in particular their influence on the primordial spatial distribution of both cores and stars is poorly understood. \citet{Myers14} find high surface density ratios for the most massive stars in their magneto-hydrodynamic simulations of star formation that include feedback. Their interpretation is that the magnetic fields are responsible for the different spatial distribution of the most massive stars. Given that the observed cores in the NGC\,2023/NGC\,2024 region of Orion B cannot have become mass segregated due to dynamics or competitive accretion, further investigation into the role of magnetic fields in this process would be highly desirable.

\section{Conclusions}

We quantify the spatial distributions of dense cores in three sub-regions of the Orion~B star-forming region, namely L1622, NGC\,2068/NGC\,2071 and NGC\,2023/NGC\,2024, using data from \citet{Kirk16a}. We determine the amount of substructure using the \citet{Cartwright04} $\mathcal{Q}$-parameter, the amount of mass segregation using the \citet{Allison09a} $\Lambda_{\rm MSR}$ ratio, the relative surface density of the most massive cores using the \citet{Maschberger11} $\Sigma_{\rm LDR}$ technique and the relative depth of the gravitational potential around the most massive cores, $\Phi_{\rm PDR}$ \citep{Parker17b}. Our conclusions are the following:

(i) In contrast to \citet{Kirk16b}, who calculated $\mathcal{Q}$-parameters consistent with smooth or centrally concentrated distributions, we find $\mathcal{Q} < 0.8$ for all three regions, which suggests a substructured or hierarchical distribution. We attribute the high values found by \citet{Kirk16b} to a flaw in their normalisation method, which uses a convex hull area instead of the area of a circle (see Appendix).

(ii) The dense cores in L1622 are not mass segregated, but the cores in NGC\,2068/NGC\,2071 are mildly mass segregated ($\Lambda_{\rm MSR} \sim 2$ for the forty most massive cores). NGC\,2023/NGC\,2024 is significantly mass segregated ($\Lambda_{\rm MSR} = 28$ for the four most massive cores, and $\Lambda_{\rm MSR} = 3.9$ for the ten most massive cores).

(iii) The most massive cores in  NGC\,2068/NGC\,2071 and NGC\,2023/NGC\,2024 lie in areas of relatively high local surface density, as well as sitting in a deeper gravitational potential than the lower-mass stars. 

(iv) Given the degree of spatial substructure in all three regions, the difference in the spatial distributions of the most massive cores compared to lower-mass cores (assuming observational biases are not wholly responsible) cannot be attributed to dynamical evolution of the cores. Instead, the observed distributions must reflect the outcome of the star formation process.

(v) The presence of primordial mass segregation in the dense cores does not necessarily support the competitive accretion theory of star formation, as hydrodynamical simulations where this process dominates do not always display differences in the spatial distributions of the most massive stars, especially in regions with high external feedback \citep{Parker17b}, such as Orion~B \citep{Pety17}.

(vi) Differences in the spatial distributions of massive cores (and stars) have been attributed to the presence of magentic fields \citep[e.g.][]{Myers14}. This idea warrants further investigation as it specifically predicts a different spatial distribution for the most massive cores/stars, even in the prescence of strong feedback, which appears to be the case in Orion~B. Simulations that do not include magnetic fields, but do include feedback have shown that the most massive stars do not attain a different spatial distribution to lower-mass objects \citep{Parker15a}.

In future papers we will investigate the spatial distributions of prestellar cores in other star-forming regions, as well as in hydrodynamical simulations of star-formation.

\section*{Acknowledgments}

I am grateful to the referee, Olly Lomax, whose comments and suggestions improved the original manuscript. I acknowledge support from the Royal Society in the form of a Dorothy Hodgkin Fellowship. I wish to thank the midwives at Sheffield's Jessop Wing maternity ward for safely delivering my son, Ilya James Gladwin, during the writing of this paper.

\bibliography{general_ref}

\begin{thebibliography}{}
\makeatletter
\relax
\def\mn@urlcharsother{\let\do\@makeother \do\$\do\&\do\#\do\^\do\_\do\%\do\~}
\def\mn@doi{\begingroup\mn@urlcharsother \@ifnextchar [ {\mn@doi@}
  {\mn@doi@[]}}
\def\mn@doi@[#1]#2{\def\@tempa{#1}\ifx\@tempa\@empty \href
  {http://dx.doi.org/#2} {doi:#2}\else \href {http://dx.doi.org/#2} {#1}\fi
  \endgroup}
\def\mn@eprint#1#2{\mn@eprint@#1:#2::\@nil}
\def\mn@eprint@arXiv#1{\href {http://arxiv.org/abs/#1} {{\tt arXiv:#1}}}
\def\mn@eprint@dblp#1{\href {http://dblp.uni-trier.de/rec/bibtex/#1.xml}
  {dblp:#1}}
\def\mn@eprint@#1:#2:#3:#4\@nil{\def\@tempa {#1}\def\@tempb {#2}\def\@tempc
  {#3}\ifx \@tempc \@empty \let \@tempc \@tempb \let \@tempb \@tempa \fi \ifx
  \@tempb \@empty \def\@tempb {arXiv}\fi \@ifundefined
  {mn@eprint@\@tempb}{\@tempb:\@tempc}{\expandafter \expandafter \csname
  mn@eprint@\@tempb\endcsname \expandafter{\@tempc}}}

\bibitem[\protect\citeauthoryear{Adams, Hollenbach, Laughlin  \& Gorti}{Adams
  et~al.}{2004}]{Adams04}
Adams F.~C.,  Hollenbach D.,  Laughlin G.,   Gorti U.,  2004, ApJ, 611, 360

\bibitem[\protect\citeauthoryear{{Alfaro} \& {Gonz{\'a}lez}}{{Alfaro} \&
  {Gonz{\'a}lez}}{2016}]{Alfaro16}
{Alfaro} E.~J.,  {Gonz{\'a}lez} M.,  2016, \mn@doi [MNRAS]
  {10.1093/mnras/stv2822}, \href
  {http://adsabs.harvard.edu/abs/2016MNRAS.456.2900A} {456, 2900}

\bibitem[\protect\citeauthoryear{Allison, Goodwin, Parker, {Portegies Zwart},
  de Grijs  \& Kouwenhoven}{Allison et~al.}{2009}]{Allison09a}
Allison R.~J.,  Goodwin S.~P.,  Parker R.~J.,  {Portegies Zwart} S.~F.,  de
  Grijs R.,   Kouwenhoven M. B.~N.,  2009, MNRAS, 395, 1449

\bibitem[\protect\citeauthoryear{Allison, Goodwin, Parker, {Portegies Zwart}
  \& de Grijs}{Allison et~al.}{2010}]{Allison10}
Allison R.~J.,  Goodwin S.~P.,  Parker R.~J.,  {Portegies Zwart} S.~F.,   de
  Grijs R.,  2010, MNRAS, 407, 1098

\bibitem[\protect\citeauthoryear{{Alves}, {Lombardi}  \& {Lada}}{{Alves}
  et~al.}{2007}]{Alves07}
{Alves} J.,  {Lombardi} M.,   {Lada} C.~J.,  2007, \mn@doi [\aap]
  {10.1051/0004-6361:20066389}, \href
  {http://adsabs.harvard.edu/abs/2007A%26A...462L..17A} {462, L17}

\bibitem[\protect\citeauthoryear{{Andr{\'e}} et~al.,}{{Andr{\'e}}
  et~al.}{2010}]{Andre10}
{Andr{\'e}} P.,  et~al., 2010, \mn@doi [A\&A] {10.1051/0004-6361/201014666},
  518, L102

\bibitem[\protect\citeauthoryear{{Ascenso}, {Alves}  \& {Lago}}{{Ascenso}
  et~al.}{2009}]{Ascenso09}
{Ascenso} J.,  {Alves} J.,   {Lago} M.~T.~V.~T.,  2009, \mn@doi [A\&A]
  {10.1051/0004-6361/200809886}, 495, 147

\bibitem[\protect\citeauthoryear{Bastian, Gieles, Ercolano  \&
  Gutermuth}{Bastian et~al.}{2009}]{Bastian09}
Bastian N.,  Gieles M.,  Ercolano B.,   Gutermuth R.,  2009, MNRAS, 392, 868

\bibitem[\protect\citeauthoryear{Bate \& Bonnell}{Bate \&
  Bonnell}{2005}]{Bate05}
Bate M.~R.,  Bonnell I.~A.,  2005, MNRAS, 356, 1201

\bibitem[\protect\citeauthoryear{Bate, Clarke  \& McCaughrean}{Bate
  et~al.}{1998}]{Bate98b}
Bate M.~R.,  Clarke C.~J.,   McCaughrean M.~J.,  1998, MNRAS, 297, 1163

\bibitem[\protect\citeauthoryear{Bonnell \& Davies}{Bonnell \&
  Davies}{1998}]{Bonnell98}
Bonnell I.~A.,  Davies M.~B.,  1998, MNRAS, 295, 691

\bibitem[\protect\citeauthoryear{{Bonnell}, {Bate}  \& {Zinnecker}}{{Bonnell}
  et~al.}{1998}]{Bonnell98b}
{Bonnell} I.~A.,  {Bate} M.~R.,   {Zinnecker} H.,  1998, \mn@doi [MNRAS]
  {10.1046/j.1365-8711.1998.01590.x}, \href
  {http://adsabs.harvard.edu/abs/1998MNRAS.298...93B} {298, 93}

\bibitem[\protect\citeauthoryear{Bonnell, Bate, Clarke  \& Pringle}{Bonnell
  et~al.}{2001}]{Bonnell01}
Bonnell I.~A.,  Bate M.~R.,  Clarke C.~J.,   Pringle J.~E.,  2001, MNRAS, 323,
  785

\bibitem[\protect\citeauthoryear{Bonnell, Clark  \& Bate}{Bonnell
  et~al.}{2008}]{Bonnell08}
Bonnell I.~A.,  Clark P.~C.,   Bate M.~R.,  2008, MNRAS, 389, 1556

\bibitem[\protect\citeauthoryear{Bressert et~al.,}{Bressert
  et~al.}{2010}]{Bressert10}
Bressert E.,  et~al., 2010, MNRAS, 409, L54

\bibitem[\protect\citeauthoryear{Cartwright}{Cartwright}{2009}]{Cartwright09b}
Cartwright A.,  2009, MNRAS, 400, 1427

\bibitem[\protect\citeauthoryear{Cartwright \& Whitworth}{Cartwright \&
  Whitworth}{2004}]{Cartwright04}
Cartwright A.,  Whitworth A.~P.,  2004, MNRAS, 348, 589

\bibitem[\protect\citeauthoryear{{Cartwright} \& {Whitworth}}{{Cartwright} \&
  {Whitworth}}{2009}]{Cartwright09a}
{Cartwright} A.,  {Whitworth} A.~P.,  2009, \mn@doi [MNRAS]
  {10.1111/j.1365-2966.2008.14055.x}, \href
  {http://adsabs.harvard.edu/abs/2009MNRAS.392..341C} {392, 341}

\bibitem[\protect\citeauthoryear{{Cartwright}, {Whitworth}  \&
  {Nutter}}{{Cartwright} et~al.}{2006}]{Cartwright06}
{Cartwright} A.,  {Whitworth} A.~P.,   {Nutter} D.,  2006, \mn@doi [MNRAS]
  {10.1111/j.1365-2966.2006.10389.x}, \href
  {http://adsabs.harvard.edu/abs/2006MNRAS.369.1411C} {369, 1411}

\bibitem[\protect\citeauthoryear{Casertano \& Hut}{Casertano \&
  Hut}{1985}]{Casertano85}
Casertano S.,  Hut P.,  1985, ApJ, 298, 80

\bibitem[\protect\citeauthoryear{{Delgado}, {Djupvik}, {Costado}  \&
  {Alfaro}}{{Delgado} et~al.}{2013}]{Delgado13}
{Delgado} A.~J.,  {Djupvik} A.~A.,  {Costado} M.~T.,   {Alfaro} E.~J.,  2013,
  MNRAS, \href {http://adsabs.harvard.edu/abs/2013arXiv1307.4290J} {435, 429}

\bibitem[\protect\citeauthoryear{{Dib}, {Schmeja}  \& {Parker}}{{Dib}
  et~al.}{2018}]{Dib18}
{Dib} S.,  {Schmeja} S.,   {Parker} R.~J.,  2018, \mn@doi [\mnras]
  {10.1093/mnras/stx2413}, \href
  {http://adsabs.harvard.edu/abs/2018MNRAS.473..849D} {473, 849}

\bibitem[\protect\citeauthoryear{{Dom{\'{\i}}nguez}, {Fellhauer}, {Bla{\~n}a},
  {Farias}  \& {Dabringhausen}}{{Dom{\'{\i}}nguez} et~al.}{2017}]{Dominguez17}
{Dom{\'{\i}}nguez} R.,  {Fellhauer} M.,  {Bla{\~n}a} M.,  {Farias} J.~P.,
  {Dabringhausen} J.,  2017, \mn@doi [\mnras] {10.1093/mnras/stx1883}, \href
  {http://adsabs.harvard.edu/abs/2017MNRAS.472..465D} {472, 465}

\bibitem[\protect\citeauthoryear{{Elmegreen}}{{Elmegreen}}{2002}]{Elmegreen02}
{Elmegreen} B.~G.,  2002, \mn@doi [\apj] {10.1086/324384}, \href
  {http://adsabs.harvard.edu/abs/2002ApJ...564..773E} {564, 773}

\bibitem[\protect\citeauthoryear{{Elmegreen} \& {Falgarone}}{{Elmegreen} \&
  {Falgarone}}{1996}]{Elmegreen96}
{Elmegreen} B.~G.,  {Falgarone} E.,  1996, \mn@doi [\apj] {10.1086/178009},
  \href {http://adsabs.harvard.edu/abs/1996ApJ...471..816E} {471, 816}

\bibitem[\protect\citeauthoryear{{Elmegreen}, {Hurst}  \& {Koenig}}{{Elmegreen}
  et~al.}{2014}]{Elmegreen14}
{Elmegreen} B.~G.,  {Hurst} R.,   {Koenig} X.,  2014, \mn@doi [\apjl]
  {10.1088/2041-8205/782/1/L1}, \href
  {http://adsabs.harvard.edu/abs/2014ApJ...782L...1E} {782, L1}

\bibitem[\protect\citeauthoryear{{Foster} et~al.,}{{Foster}
  et~al.}{2015}]{Foster15}
{Foster} J.~B.,  et~al., 2015, \mn@doi [ApJ] {10.1088/0004-637X/799/2/136},
  \href {http://adsabs.harvard.edu/abs/2015ApJ...799..136F} {799, 136}

\bibitem[\protect\citeauthoryear{{Girichidis}, {Federrath}, {Allison},
  {Banerjee}  \& {Klessen}}{{Girichidis} et~al.}{2012}]{Girichidis12}
{Girichidis} P.,  {Federrath} C.,  {Allison} R.,  {Banerjee} R.,   {Klessen}
  R.~S.,  2012, MNRAS, 420, 3264

\bibitem[\protect\citeauthoryear{{Gonz{\'a}lez} \& {Alfaro}}{{Gonz{\'a}lez} \&
  {Alfaro}}{2017}]{Gonzalez17}
{Gonz{\'a}lez} M.,  {Alfaro} E.~J.,  2017, \mn@doi [\mnras]
  {10.1093/mnras/stw2855}, \href
  {http://adsabs.harvard.edu/abs/2017MNRAS.465.1889G} {465, 1889}

\bibitem[\protect\citeauthoryear{Goodwin \& Whitworth}{Goodwin \&
  Whitworth}{2004}]{Goodwin04a}
Goodwin S.~P.,  Whitworth A.~P.,  2004, A\&A, 413, 929

\bibitem[\protect\citeauthoryear{Goodwin, Kroupa, Goodman  \& Burkert}{Goodwin
  et~al.}{2007}]{Goodwin07}
Goodwin S.~P.,  Kroupa P.,  Goodman A.,   Burkert A.,  2007, in Reipurth B.,
  Jewitt D.,   Keil K.,  eds, {Protostars and Planets V}. pp 133--147

\bibitem[\protect\citeauthoryear{{Gouliermis}, {Hony}  \&
  {Klessen}}{{Gouliermis} et~al.}{2014}]{Gouliermis14}
{Gouliermis} D.~A.,  {Hony} S.,   {Klessen} R.~S.,  2014, \mn@doi [MNRAS]
  {10.1093/mnras/stu228}, \href
  {http://cdsads.u-strasbg.fr/abs/2014MNRAS.439.3775G} {439, 3775}

\bibitem[\protect\citeauthoryear{Gutermuth, Megeath, Myers, Allen  \&
  Fazio}{Gutermuth et~al.}{2009}]{Gutermuth09}
Gutermuth R.~A.,  Megeath S.~T.,  Myers P.~C.,  Allen L.~E.,   Fazio J. L. P.
  G.~G.,  2009, ApJS, 184, 18

\bibitem[\protect\citeauthoryear{{Habing}}{{Habing}}{1968}]{Habing68}
{Habing} H.~J.,  1968, BAIN, \href
  {http://adsabs.harvard.edu/abs/1968BAN....19..421H} {19, 421}

\bibitem[\protect\citeauthoryear{{Hacar}, {Tafalla}, {Kauffmann}  \&
  {Kov{\'a}cs}}{{Hacar} et~al.}{2013}]{Hacar13}
{Hacar} A.,  {Tafalla} M.,  {Kauffmann} J.,   {Kov{\'a}cs} A.,  2013, A\&A,
  554, A55

\bibitem[\protect\citeauthoryear{{Hatchell} \& {Fuller}}{{Hatchell} \&
  {Fuller}}{2008}]{Hatchell08}
{Hatchell} J.,  {Fuller} G.~A.,  2008, \mn@doi [\aap]
  {10.1051/0004-6361:20079213}, \href
  {http://adsabs.harvard.edu/abs/2008A%26A...482..855H} {482, 855}

\bibitem[\protect\citeauthoryear{{Henshaw} et~al.,}{{Henshaw}
  et~al.}{2016}]{Henshaw16}
{Henshaw} J.~D.,  et~al., 2016, \mn@doi [\mnras] {10.1093/mnras/stw1794}, \href
  {http://adsabs.harvard.edu/abs/2016MNRAS.463..146H} {463, 146}

\bibitem[\protect\citeauthoryear{Hillenbrand \& Hartmann}{Hillenbrand \&
  Hartmann}{1998}]{Hillenbrand98}
Hillenbrand L.~A.,  Hartmann L.~W.,  1998, ApJ, 492, 540

\bibitem[\protect\citeauthoryear{Hoyle}{Hoyle}{1953}]{Hoyle53}
Hoyle F.,  1953, ApJ, 118, 513

\bibitem[\protect\citeauthoryear{{Jaffa}, {Whitworth}  \& {Lomax}}{{Jaffa}
  et~al.}{2017}]{Jaffa17}
{Jaffa} S.~E.,  {Whitworth} A.~P.,   {Lomax} O.,  2017, \mn@doi [MNRAS]
  {10.1093/mnras/stw3140}, \href
  {http://adsabs.harvard.edu/abs/2017MNRAS.466.1082J} {466, 1082}

\bibitem[\protect\citeauthoryear{{Kainulainen}, {Lada}, {Rathborne}  \&
  {Alves}}{{Kainulainen} et~al.}{2009}]{Kainulainen09}
{Kainulainen} J.,  {Lada} C.~J.,  {Rathborne} J.~M.,   {Alves} J.~F.,  2009,
  \mn@doi [\aap] {10.1051/0004-6361/200810987}, \href
  {http://adsabs.harvard.edu/abs/2009A%26A...497..399K} {497, 399}

\bibitem[\protect\citeauthoryear{{Kainulainen}, {Stutz}, {Stanke},
  {Abreu-Vicente}, {Beuther}, {Henning}, {Johnston}  \&
  {Megeath}}{{Kainulainen} et~al.}{2017}]{Kainulainen17}
{Kainulainen} J.,  {Stutz} A.~M.,  {Stanke} T.,  {Abreu-Vicente} J.,  {Beuther}
  H.,  {Henning} T.,  {Johnston} K.~G.,   {Megeath} S.~T.,  2017, \mn@doi
  [\aap] {10.1051/0004-6361/201628481}, \href
  {http://adsabs.harvard.edu/abs/2017A%26A...600A.141K} {600, A141}

\bibitem[\protect\citeauthoryear{{Kauffmann}, {Pillai}  \&
  {Goldsmith}}{{Kauffmann} et~al.}{2013}]{Kauffmann13}
{Kauffmann} J.,  {Pillai} T.,   {Goldsmith} P.~F.,  2013, \mn@doi [ApJ]
  {10.1088/0004-637X/779/2/185}, \href
  {http://adsabs.harvard.edu/abs/2013ApJ...779..185K} {779, 185}

\bibitem[\protect\citeauthoryear{Kirk \& Myers}{Kirk \& Myers}{2011}]{Kirk10}
Kirk H.,  Myers P.~C.,  2011, ApJ, 727, 64

\bibitem[\protect\citeauthoryear{{Kirk}, {Offner}  \& {Redmond}}{{Kirk}
  et~al.}{2014}]{Kirk14}
{Kirk} H.,  {Offner} S.~S.~R.,   {Redmond} K.~J.,  2014, \mn@doi [MNRAS]
  {10.1093/mnras/stu052}, \href
  {http://adsabs.harvard.edu/abs/2014MNRAS.439.1765K} {439, 1765}

\bibitem[\protect\citeauthoryear{{Kirk} et~al.,}{{Kirk}
  et~al.}{2016a}]{Kirk16a}
{Kirk} H.,  et~al., 2016a, \mn@doi [\apj] {10.3847/0004-637X/817/2/167}, \href
  {http://adsabs.harvard.edu/abs/2016ApJ...817..167K} {817, 167}

\bibitem[\protect\citeauthoryear{{Kirk} et~al.,}{{Kirk}
  et~al.}{2016b}]{Kirk16b}
{Kirk} H.,  et~al., 2016b, \mn@doi [\apj] {10.3847/0004-637X/821/2/98}, \href
  {http://adsabs.harvard.edu/abs/2016ApJ...821...98K} {821, 98}

\bibitem[\protect\citeauthoryear{{K{\"o}nyves} et~al.,}{{K{\"o}nyves}
  et~al.}{2010}]{Konyves10}
{K{\"o}nyves} V.,  et~al., 2010, \mn@doi [\aap] {10.1051/0004-6361/201014689},
  \href {http://adsabs.harvard.edu/abs/2010A%26A...518L.106K} {518, L106}

\bibitem[\protect\citeauthoryear{Kroupa}{Kroupa}{1995}]{Kroupa95a}
Kroupa P.,  1995, MNRAS, 277, 1491

\bibitem[\protect\citeauthoryear{Kruijssen, Maschberger, Moeckel, Clarke,
  Bastian  \& Bonnell}{Kruijssen et~al.}{2012}]{Kruijssen12a}
Kruijssen J. M.~D.,  Maschberger T.,  Moeckel N.,  Clarke C.~J.,  Bastian N.,
  Bonnell I.~A.,  2012, MNRAS, 419, 841

\bibitem[\protect\citeauthoryear{{Kuhn}, {Getman}, {Feigelson}, {Sills},
  {Gromadzki}, {Medina}, {Borissova}  \& {Kurtev}}{{Kuhn}
  et~al.}{2017}]{Kuhn17}
{Kuhn} M.~A.,  {Getman} K.~V.,  {Feigelson} E.~D.,  {Sills} A.,  {Gromadzki}
  M.,  {Medina} N.,  {Borissova} J.,   {Kurtev} R.,  2017, \mn@doi [\aj]
  {10.3847/1538-3881/aa9177}, \href
  {http://adsabs.harvard.edu/abs/2017AJ....154..214K} {154, 214}

\bibitem[\protect\citeauthoryear{{K{\"u}pper}, {Maschberger}, {Kroupa}  \&
  {Baumgardt}}{{K{\"u}pper} et~al.}{2011}]{Kupper11}
{K{\"u}pper} A.~H.~W.,  {Maschberger} T.,  {Kroupa} P.,   {Baumgardt} H.,
  2011, \mn@doi [MNRAS] {10.1111/j.1365-2966.2011.19412.x}, \href
  {http://adsabs.harvard.edu/abs/2011MNRAS.417.2300K} {417, 2300}

\bibitem[\protect\citeauthoryear{{Kuznetsova}, {Hartmann}  \&
  {Ballesteros-Paredes}}{{Kuznetsova} et~al.}{2015}]{Kuznetsova15}
{Kuznetsova} A.,  {Hartmann} L.,   {Ballesteros-Paredes} J.,  2015, \mn@doi
  [ApJ] {10.1088/0004-637X/815/1/27}, \href
  {http://adsabs.harvard.edu/abs/2015ApJ...815...27K} {815, 27}

\bibitem[\protect\citeauthoryear{Lada \& Lada}{Lada \& Lada}{2003}]{Lada03}
Lada C.~J.,  Lada E.~A.,  2003, ARA\&A, 41, 57

\bibitem[\protect\citeauthoryear{Littlefair, Naylor, Harries, Retter  \&
  {O'Toole}}{Littlefair et~al.}{2004}]{Littlefair04}
Littlefair S.~P.,  Naylor T.,  Harries T.~J.,  Retter A.,   {O'Toole} S.,
  2004, MNRAS, 347, 937

\bibitem[\protect\citeauthoryear{{Lomax}, {Whitworth}  \& {Cartwright}}{{Lomax}
  et~al.}{2011}]{Lomax11}
{Lomax} O.,  {Whitworth} A.~P.,   {Cartwright} A.,  2011, \mn@doi [MNRAS]
  {10.1111/j.1365-2966.2010.17935.x}, \href
  {http://adsabs.harvard.edu/abs/2011MNRAS.412..627L} {412, 627}

\bibitem[\protect\citeauthoryear{{Lomax}, {Whitworth}, {Hubber}, {Stamatellos}
  \& {Walch}}{{Lomax} et~al.}{2014}]{Lomax14}
{Lomax} O.,  {Whitworth} A.~P.,  {Hubber} D.~A.,  {Stamatellos} D.,   {Walch}
  S.,  2014, \mn@doi [\mnras] {10.1093/mnras/stu177}, \href
  {http://adsabs.harvard.edu/abs/2014MNRAS.439.3039L} {439, 3039}

\bibitem[\protect\citeauthoryear{Maschberger \& Clarke}{Maschberger \&
  Clarke}{2011}]{Maschberger11}
Maschberger T.,  Clarke C.~J.,  2011, MNRAS, 416, 541

\bibitem[\protect\citeauthoryear{{McMillan}, {Vesperini}  \& {Portegies
  Zwart}}{{McMillan} et~al.}{2007}]{McMillan07}
{McMillan} S.~L.~W.,  {Vesperini} E.,   {Portegies Zwart} S.~F.,  2007, \mn@doi
  [ApJL] {10.1086/511763}, \href
  {http://adsabs.harvard.edu/abs/2007ApJ...655L..45M} {655, L45}

\bibitem[\protect\citeauthoryear{{Moeckel} \& {Bonnell}}{{Moeckel} \&
  {Bonnell}}{2009a}]{Moeckel09a}
{Moeckel} N.,  {Bonnell} I.~A.,  2009a, \mn@doi [MNRAS]
  {10.1111/j.1365-2966.2009.14813.x}, \href
  {http://adsabs.harvard.edu/abs/2009MNRAS.396.1864M} {396, 1864}

\bibitem[\protect\citeauthoryear{{Moeckel} \& {Bonnell}}{{Moeckel} \&
  {Bonnell}}{2009b}]{Moeckel09b}
{Moeckel} N.,  {Bonnell} I.~A.,  2009b, \mn@doi [MNRAS]
  {10.1111/j.1365-2966.2009.15499.x}, \href
  {http://adsabs.harvard.edu/abs/2009MNRAS.400..657M} {400, 657}

\bibitem[\protect\citeauthoryear{{Myers}, {Klein}, {Krumholz}  \&
  {McKee}}{{Myers} et~al.}{2014}]{Myers14}
{Myers} A.~T.,  {Klein} R.~I.,  {Krumholz} M.~R.,   {McKee} C.~F.,  2014,
  \mn@doi [MNRAS] {10.1093/mnras/stu190}, \href
  {http://adsabs.harvard.edu/abs/2014MNRAS.439.3420M} {439, 3420}

\bibitem[\protect\citeauthoryear{{Offner}, {Clark}, {Hennebelle}, {Bastian},
  {Bate}, {Hopkins}, {Moraux}  \& {Whitworth}}{{Offner}
  et~al.}{2014}]{Offner14}
{Offner} S.~S.~R.,  {Clark} P.~C.,  {Hennebelle} P.,  {Bastian} N.,  {Bate}
  M.~R.,  {Hopkins} P.~F.,  {Moraux} E.,   {Whitworth} A.~P.,  2014, \mn@doi
  [Protostars and Planets VI] {10.2458/azu_uapress_9780816531240-ch003}, \href
  {http://adsabs.harvard.edu/abs/2014prpl.conf...53O} {pp 53--75}

\bibitem[\protect\citeauthoryear{Olczak, Spurzem  \& Henning}{Olczak
  et~al.}{2011}]{Olczak11}
Olczak C.,  Spurzem R.,   Henning T.,  2011, A\&A, 532, 119

\bibitem[\protect\citeauthoryear{Parker \& {Alves de Oliveira}}{Parker \&
  {Alves de Oliveira}}{2017}]{Parker17a}
Parker R.~J.,  {Alves de Oliveira} C.,  2017, MNRAS, 468, 4340

\bibitem[\protect\citeauthoryear{Parker \& Dale}{Parker \&
  Dale}{2015}]{Parker15c}
Parker R.~J.,  Dale J.~E.,  2015, MNRAS, 451, 3664

\bibitem[\protect\citeauthoryear{Parker \& Dale}{Parker \&
  Dale}{2017}]{Parker17b}
Parker R.~J.,  Dale J.~E.,  2017, MNRAS, 470, 390

\bibitem[\protect\citeauthoryear{Parker \& Goodwin}{Parker \&
  Goodwin}{2015}]{Parker15b}
Parker R.~J.,  Goodwin S.~P.,  2015, MNRAS, 449, 3381

\bibitem[\protect\citeauthoryear{Parker \& Meyer}{Parker \&
  Meyer}{2012}]{Parker12d}
Parker R.~J.,  Meyer M.~R.,  2012, MNRAS, 427, 637

\bibitem[\protect\citeauthoryear{Parker, Bouvier, Goodwin, Moraux, Allison,
  Guieu  \& G{\"u}del}{Parker et~al.}{2011a}]{Parker11b}
Parker R.~J.,  Bouvier J.,  Goodwin S.~P.,  Moraux E.,  Allison R.~J.,  Guieu
  S.,   G{\"u}del M.,  2011a, MNRAS, 412, 2489

\bibitem[\protect\citeauthoryear{Parker, Goodwin  \& Allison}{Parker
  et~al.}{2011b}]{Parker11c}
Parker R.~J.,  Goodwin S.~P.,   Allison R.~J.,  2011b, MNRAS, 418, 2565

\bibitem[\protect\citeauthoryear{Parker, Wright, Goodwin  \& Meyer}{Parker
  et~al.}{2014}]{Parker14b}
Parker R.~J.,  Wright N.~J.,  Goodwin S.~P.,   Meyer M.~R.,  2014, MNRAS, 438,
  620

\bibitem[\protect\citeauthoryear{Parker, Dale  \& Ercolano}{Parker
  et~al.}{2015}]{Parker15a}
Parker R.~J.,  Dale J.~E.,   Ercolano B.,  2015, MNRAS, 446, 4278

\bibitem[\protect\citeauthoryear{{Peretto}, {Andr{\'e}}  \&
  {Belloche}}{{Peretto} et~al.}{2006}]{Peretto06}
{Peretto} N.,  {Andr{\'e}} P.,   {Belloche} A.,  2006, \mn@doi [A\&A]
  {10.1051/0004-6361:20053324}, \href
  {http://adsabs.harvard.edu/abs/2006A\%26A...445..979P} {445, 979}

\bibitem[\protect\citeauthoryear{{Pety} et~al.,}{{Pety} et~al.}{2017}]{Pety17}
{Pety} J.,  et~al., 2017, \mn@doi [\aap] {10.1051/0004-6361/201629862}, \href
  {http://adsabs.harvard.edu/abs/2017A%26A...599A..98P} {599, A98}

\bibitem[\protect\citeauthoryear{{Porras}, {Christopher}, {Allen}, {Di
  Francesco}, {Megeath}  \& {Myers}}{{Porras} et~al.}{2003}]{Porras03}
{Porras} A.,  {Christopher} M.,  {Allen} L.,  {Di Francesco} J.,  {Megeath}
  S.~T.,   {Myers} P.~C.,  2003, \mn@doi [AJ] {10.1086/377623}, \href
  {http://adsabs.harvard.edu/abs/2003AJ....126.1916P} {126, 1916}

\bibitem[\protect\citeauthoryear{{Portegies Zwart}}{{Portegies
  Zwart}}{2016}]{Zwart16}
{Portegies Zwart} S.~F.,  2016, \mn@doi [MNRAS] {10.1093/mnras/stv2831}, \href
  {http://adsabs.harvard.edu/abs/2016MNRAS.457..313P} {457, 313}

\bibitem[\protect\citeauthoryear{{Raboud} \& {Mermilliod}}{{Raboud} \&
  {Mermilliod}}{1998}]{Raboud98}
{Raboud} D.,  {Mermilliod} J.-C.,  1998, A\&A, 333, 897

\bibitem[\protect\citeauthoryear{{Reipurth}, {Clarke}, {Boss}, {Goodwin},
  {Rodriguez}, {Stassun}, {Tokovinin}  \& {Zinnecker}}{{Reipurth}
  et~al.}{2014}]{Reipurth14}
{Reipurth} B.,  {Clarke} C.~J.,  {Boss} A.~P.,  {Goodwin} S.~P.,  {Rodriguez}
  L.~F.,  {Stassun} K.~G.,  {Tokovinin} A.,   {Zinnecker} H.,  2014, ArXiv
  e-prints: 1403.1907, \href
  {http://cdsads.u-strasbg.fr/abs/2014arXiv1403.1907R} {}

\bibitem[\protect\citeauthoryear{S{\'a}nchez \& Alfaro}{S{\'a}nchez \&
  Alfaro}{2009}]{Sanchez09}
S{\'a}nchez N.,  Alfaro E.~J.,  2009, ApJ, 696, 2086

\bibitem[\protect\citeauthoryear{Scally \& Clarke}{Scally \&
  Clarke}{2001}]{Scally01}
Scally A.,  Clarke C.,  2001, MNRAS, 325, 449

\bibitem[\protect\citeauthoryear{{Schmeja} \& {Klessen}}{{Schmeja} \&
  {Klessen}}{2006}]{Schmeja06}
{Schmeja} S.,  {Klessen} R.~S.,  2006, \mn@doi [A\&A]
  {10.1051/0004-6361:20054464}, 449, 151

\bibitem[\protect\citeauthoryear{{Schneider}, {Csengeri}, {Bontemps}, {Motte},
  {Simon}, {Hennebelle}, {Federrath}  \& {Klessen}}{{Schneider}
  et~al.}{2010}]{Schneider10}
{Schneider} N.,  {Csengeri} T.,  {Bontemps} S.,  {Motte} F.,  {Simon} R.,
  {Hennebelle} P.,  {Federrath} C.,   {Klessen} R.,  2010, \mn@doi [\aap]
  {10.1051/0004-6361/201014481}, \href
  {http://adsabs.harvard.edu/abs/2010A%26A...520A..49S} {520, A49}

\bibitem[\protect\citeauthoryear{{Smith}, {Glover}, {Klessen}  \&
  {Fuller}}{{Smith} et~al.}{2016}]{Smith16}
{Smith} R.~J.,  {Glover} S.~C.~O.,  {Klessen} R.~S.,   {Fuller} G.~A.,  2016,
  \mn@doi [\mnras] {10.1093/mnras/stv2559}, \href
  {http://adsabs.harvard.edu/abs/2016MNRAS.455.3640S} {455, 3640}

\bibitem[\protect\citeauthoryear{Wright, Parker, Goodwin  \& Drake}{Wright
  et~al.}{2014}]{Wright14}
Wright N.~J.,  Parker R.~J.,  Goodwin S.~P.,   Drake J.~J.,  2014, MNRAS, 438,
  639

\bibitem[\protect\citeauthoryear{{Zinnecker}}{{Zinnecker}}{1982}]{Zinnecker82}
{Zinnecker} H.,  1982, \mn@doi [Annals of the New York Academy of Sciences]
  {10.1111/j.1749-6632.1982.tb43399.x}, \href
  {http://adsabs.harvard.edu/abs/1982NYASA.395..226Z} {395, 226}

\bibitem[\protect\citeauthoryear{{de Grijs}, {Johnson}, {Gilmore}  \&
  {Frayn}}{{de Grijs} et~al.}{2002}]{deGrijs02}
{de Grijs} R.,  {Johnson} R.~A.,  {Gilmore} G.~F.,   {Frayn} C.~M.,  2002,
  \mn@doi [MNRAS] {10.1046/j.1365-8711.2002.05217.x}, \href
  {http://adsabs.harvard.edu/abs/2002MNRAS.331..228D} {331, 228}

\makeatother
\end{thebibliography}

\appendix

\section{Normalisation of the $\mathcal{Q}$-parameter}
\label{appendix}

\begin{figure*}
  \begin{center}
\setlength{\subfigcapskip}{10pt}
\hspace*{-1.5cm}\subfigure[$D = 1.6$]{\label{convex_hull-a}\rotatebox{270}{\includegraphics[scale=0.28]{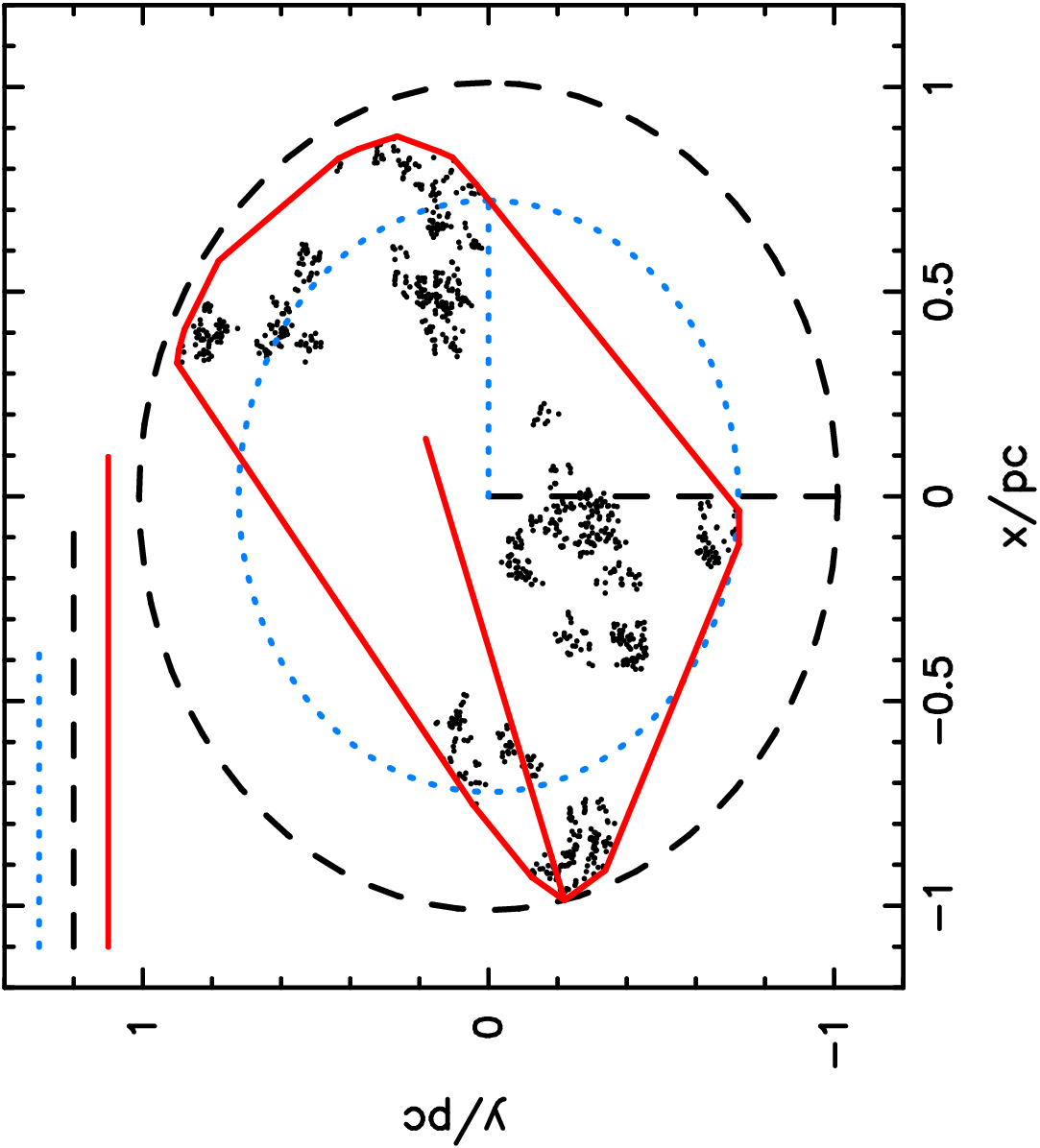}}} 
\hspace*{0.3cm} 
\subfigure[$D = 3.0$]{\label{convex_hull-b}\rotatebox{270}{\includegraphics[scale=0.28]{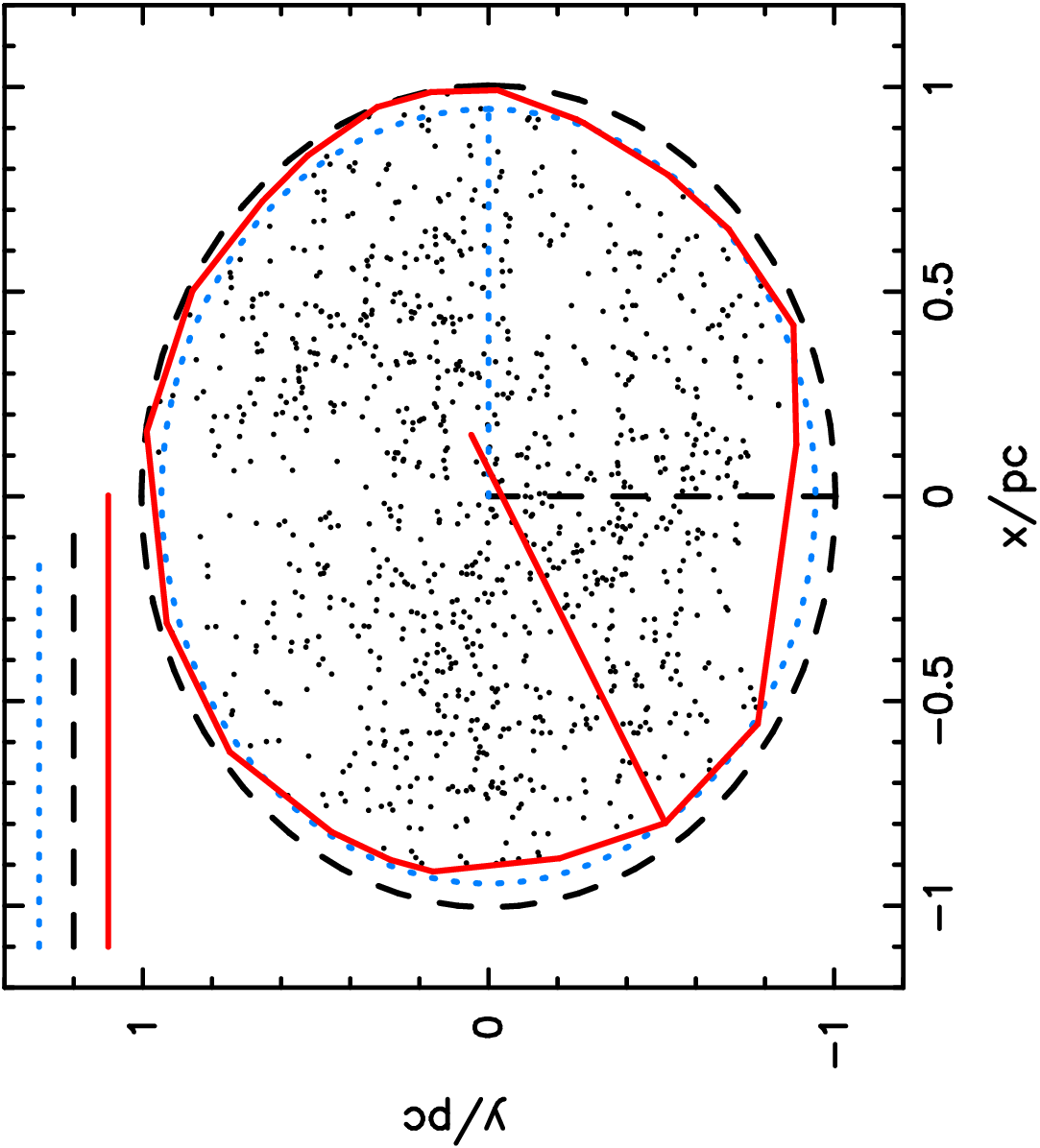}}} 
\hspace*{0.3cm}\subfigure[$n \propto r^{-2.9}$]{\label{convex_hull-c}\rotatebox{270}{\includegraphics[scale=0.28]{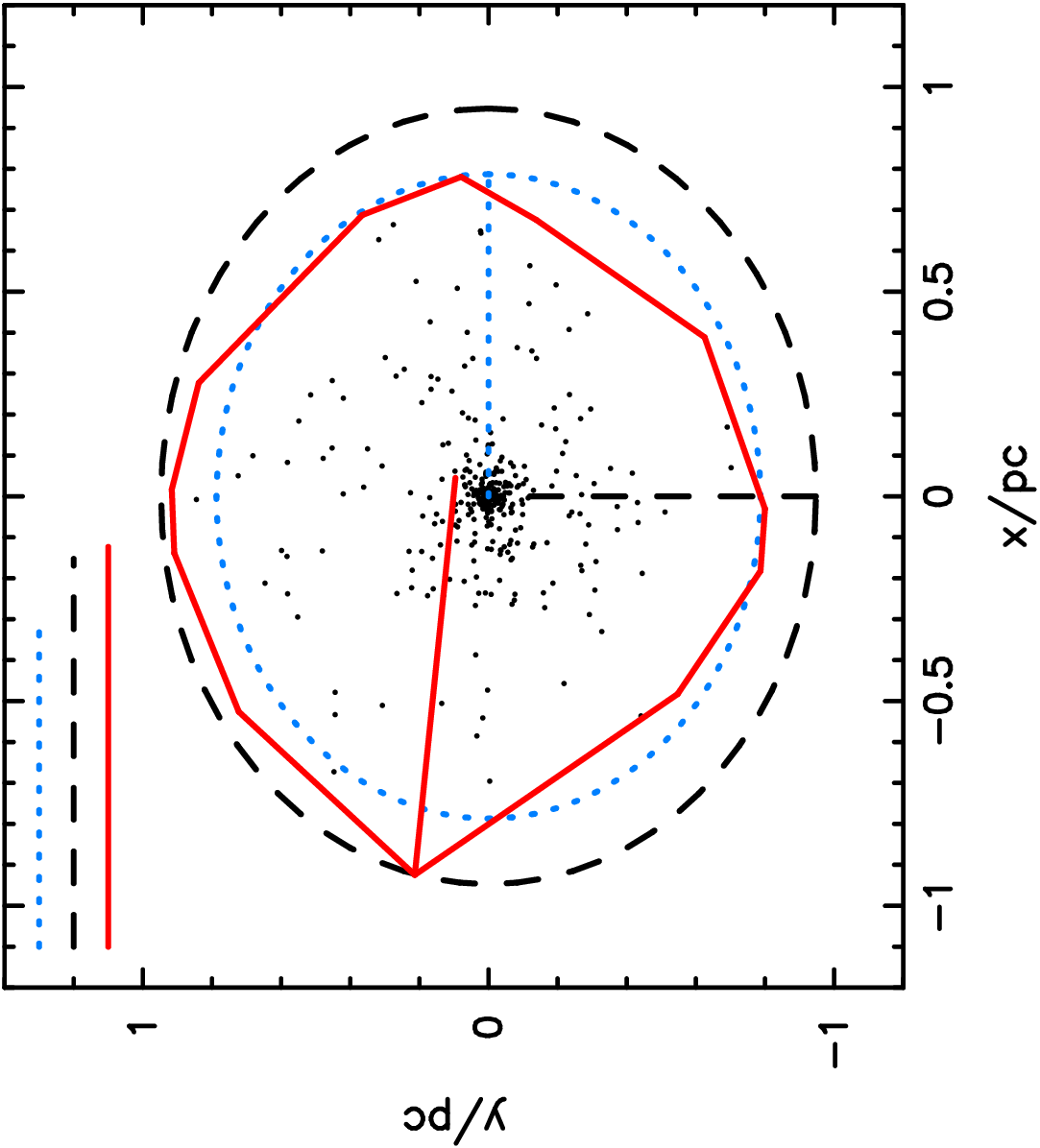}}}
\caption[bf]{Demonstration of the three methods used to normalise the $\mathcal{Q}$-parameter. The original method from \citet{Cartwright04}, where the distribution is normalised to the area $A$ of a circle with radius $R$ encompassing the most distant point is shown by the black dashed lines. The method from \citet{Kirk16b}, which uses the area of a convex hull, $A_{\rm CH}$ and a radius equal to the distance of the outermost point in the convex hull from the average position of the convex hull points, $R_{\rm CH-ex}$, is shown by the solid red lines. Finally, \citet{Schmeja06} normalise $\mathcal{Q}$ by using the area of the convex hull $A_{\rm CH}$ and drawing a circle with the radius calculated from this area, $R_{\rm CH-circ}$ (the blue dotted lines). For reference, the lengths of each of these radii are shown in the top left of each panel. We show three different geometries; a fractal with $D = 1.6$, a fractal with $D = 3.0$ and a smooth, centrally concentrated distribution with radial density profile $n \propto r^{-2.9}$.}
\label{convex_hull}
  \end{center}
\end{figure*}

\begin{figure*}
  \begin{center}
\setlength{\subfigcapskip}{10pt}
\hspace*{-1.5cm}\subfigure[Standard normalisation]{\label{qpar_test-a}\rotatebox{270}{\includegraphics[scale=0.28]{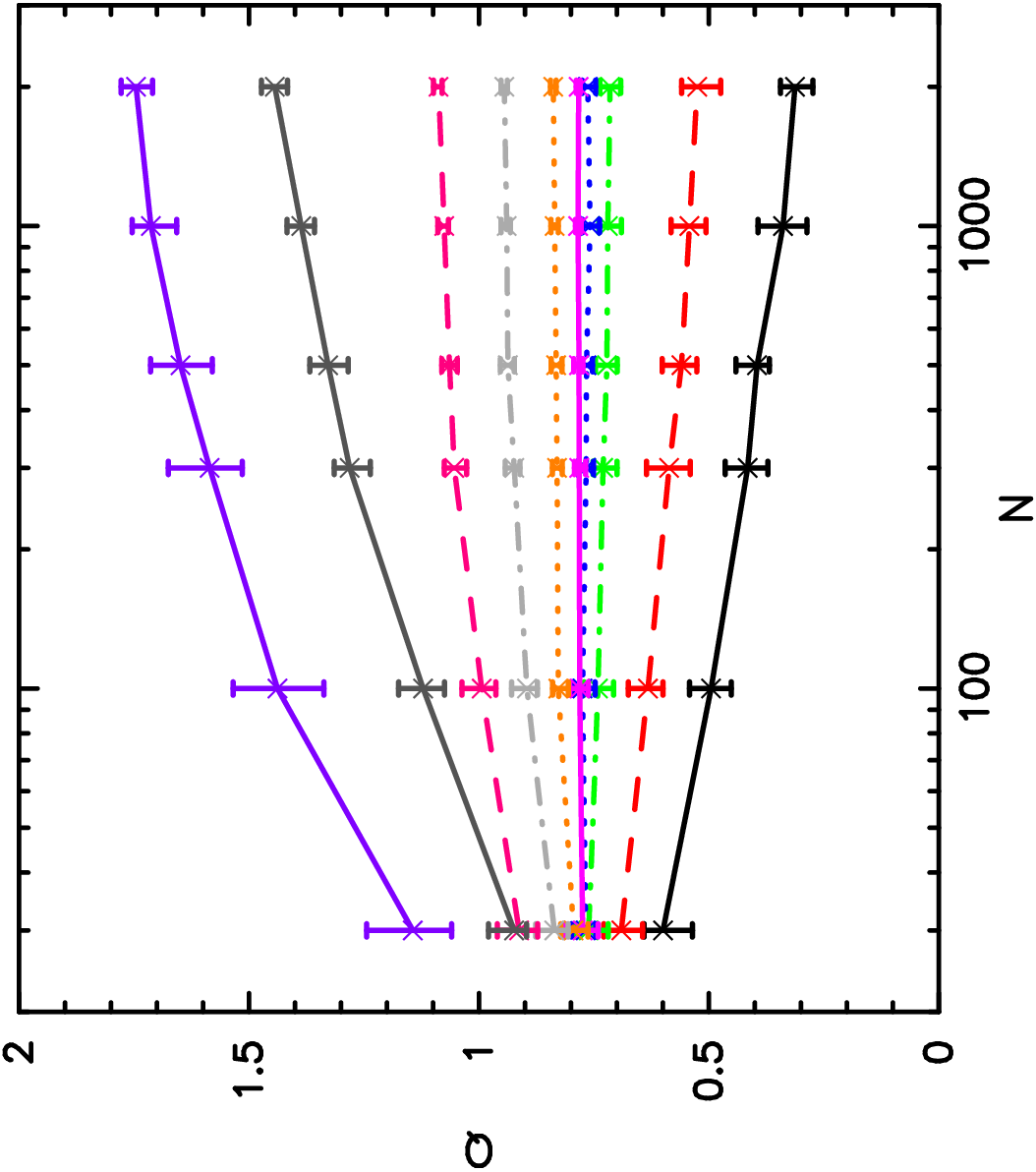}}}
\hspace*{0.3cm} 
\subfigure[Convex hull area normalisation]{\label{qpar_test-b}\rotatebox{270}{\includegraphics[scale=0.28]{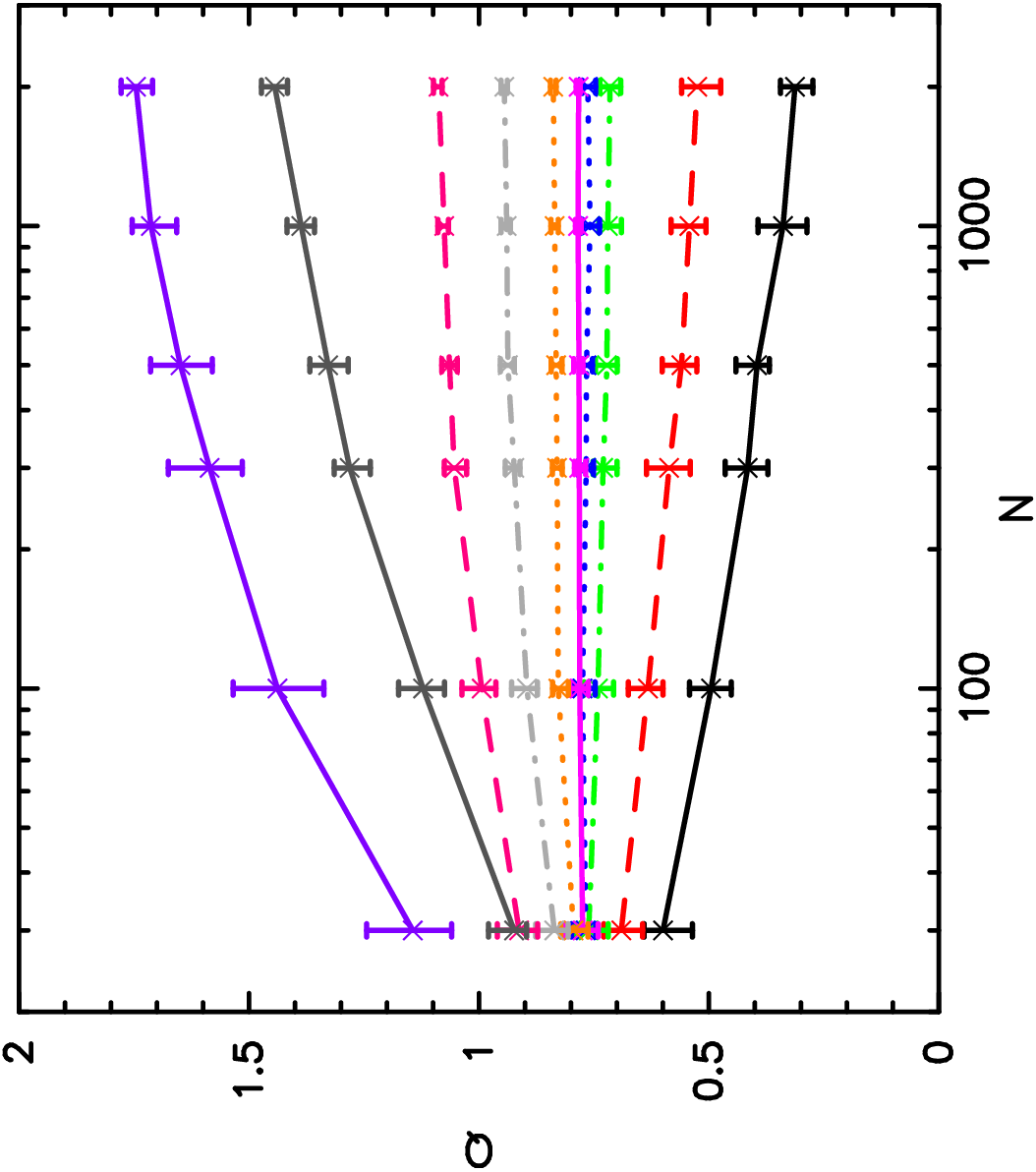}}}
\hspace*{0.3cm} 
\subfigure[Full convex hull normalisation]{\label{qpar_test-c}\rotatebox{270}{\includegraphics[scale=0.28]{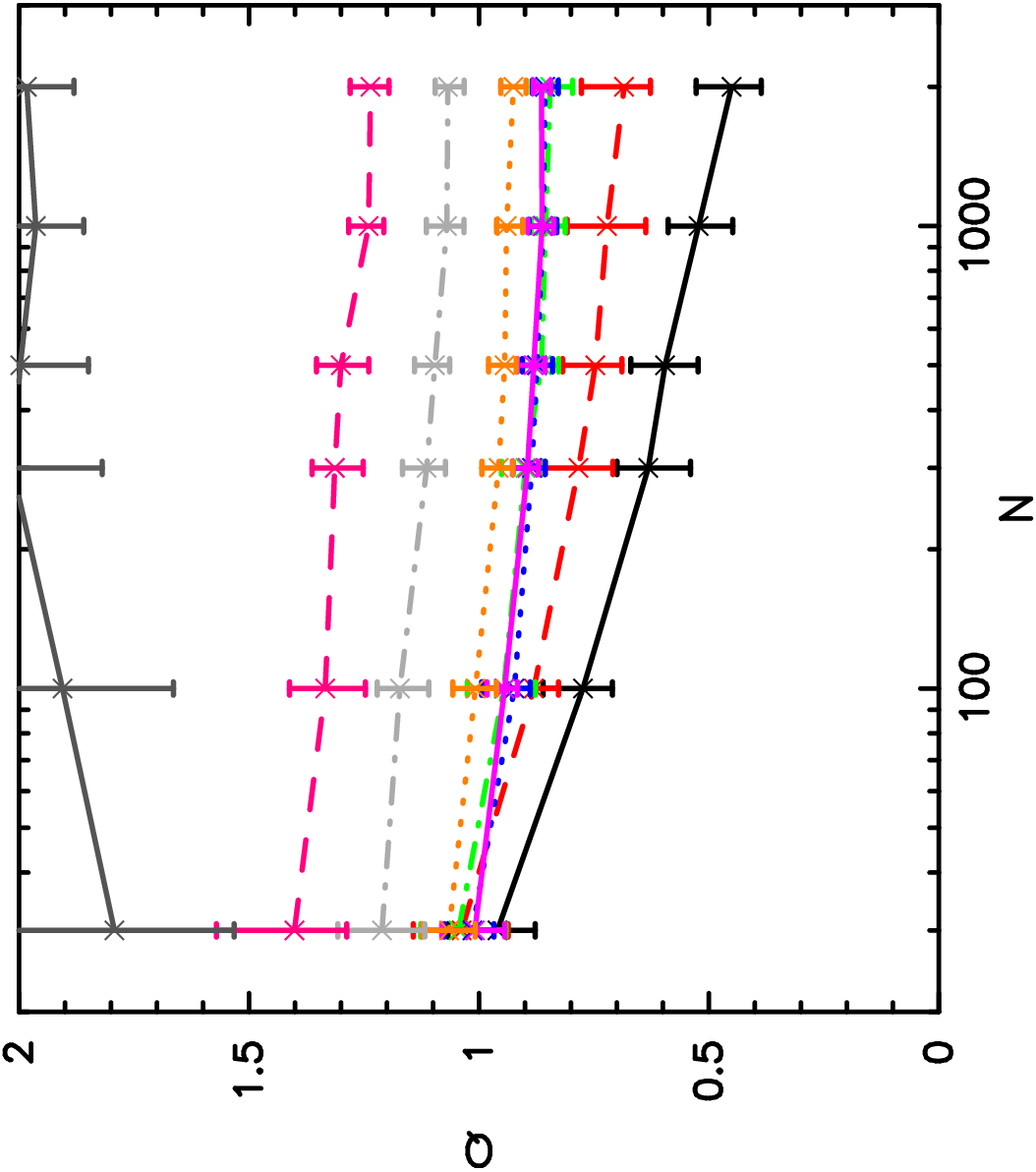}}} 
\caption[bf]{The $\mathcal{Q}$-parameter as a function of the number of objects in a distribution. In panel (a) the $\mathcal{Q}$-parameter is normalised to the area of a circle with a radius equal to the distance of the furthest point from the centre \citep{Cartwright04}. In panel (b) the  $\mathcal{Q}$-parameter is normalised to the area of a convex hull, and a radius of a circle with an area equal to that of the convex hull \citep{Schmeja06}. In panel (c) the $\mathcal{Q}$-parameter is normalised to the area of a convex hull, and a `radius' equal to the distance between the furthest point and the centre of the convex hull \citep{Kirk16b}. From bottom to top, the lines represent different morphologies, starting with a highly substructured distribution and becoming progressively smoother and more centrally concentrated. We show fractal distributions with fractal dimension $D = 1.6$ (black solid lines), $D = 2.0$ (red dashed lines), $D = 2.6$ (green dot-dashed lines), $D = 3.0$ (blue dotted lines) and smooth, centrally concentrated radial density profiles with $n \propto r^{0}$ (magenta solid lines), $n \propto r^{-1.0}$ (orange dotted lines), $n \propto r^{-2.0}$ (grey dot-dashed line), $n \propto r^{-2.5}$ (magenta dashed lines), a Plummer sphere (solid charcoal grey lines) and $n \propto r^{-2.9}$ (purple solid lines). The error bars represent the interquartile range of 100 realisations of each distribution. }
\label{qpar_test}
  \end{center}
\end{figure*}

\begin{figure*}
  \begin{center}
\setlength{\subfigcapskip}{10pt}
\hspace*{-1.5cm}\subfigure[Standard normalisation]{\label{mbar_sbar_test-a}\rotatebox{270}{\includegraphics[scale=0.28]{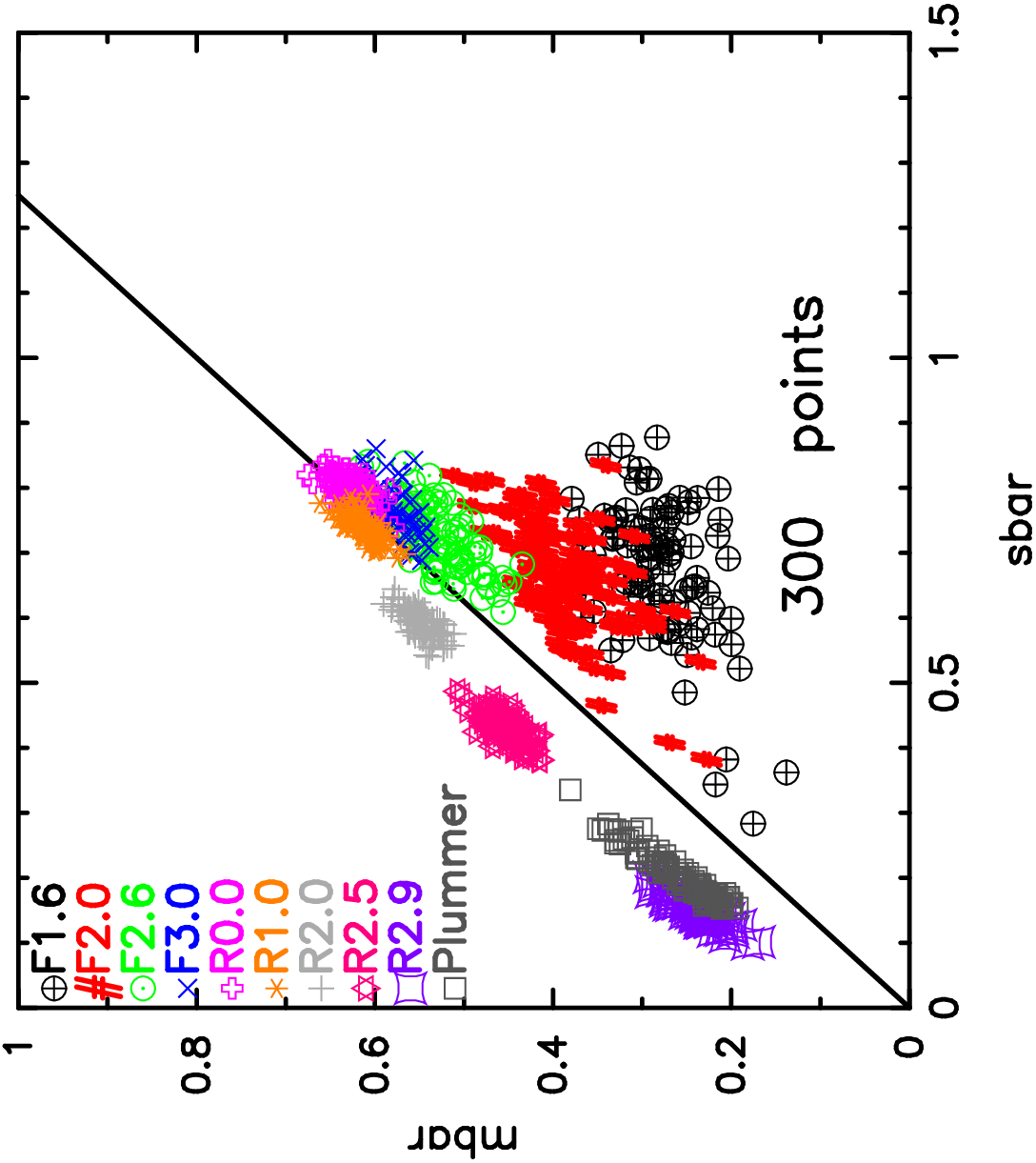}}}
\hspace*{0.3cm} 
\subfigure[Convex hull area normalisation]{\label{mbar_sbar_test-b}\rotatebox{270}{\includegraphics[scale=0.28]{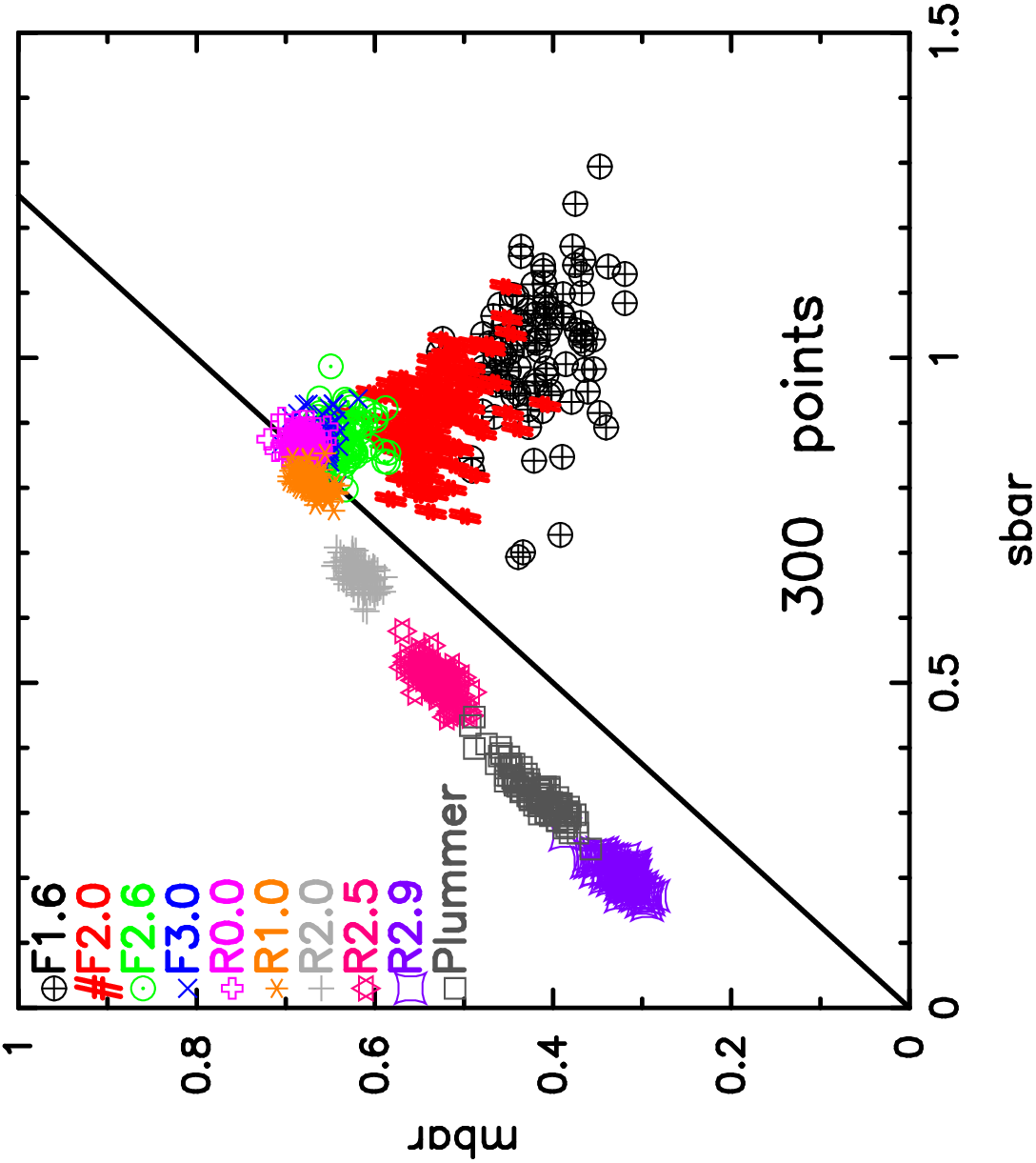}}} 
\hspace*{0.3cm}\subfigure[Full convex hull normalisation]{\label{mbar_sbar_test-c}\rotatebox{270}{\includegraphics[scale=0.28]{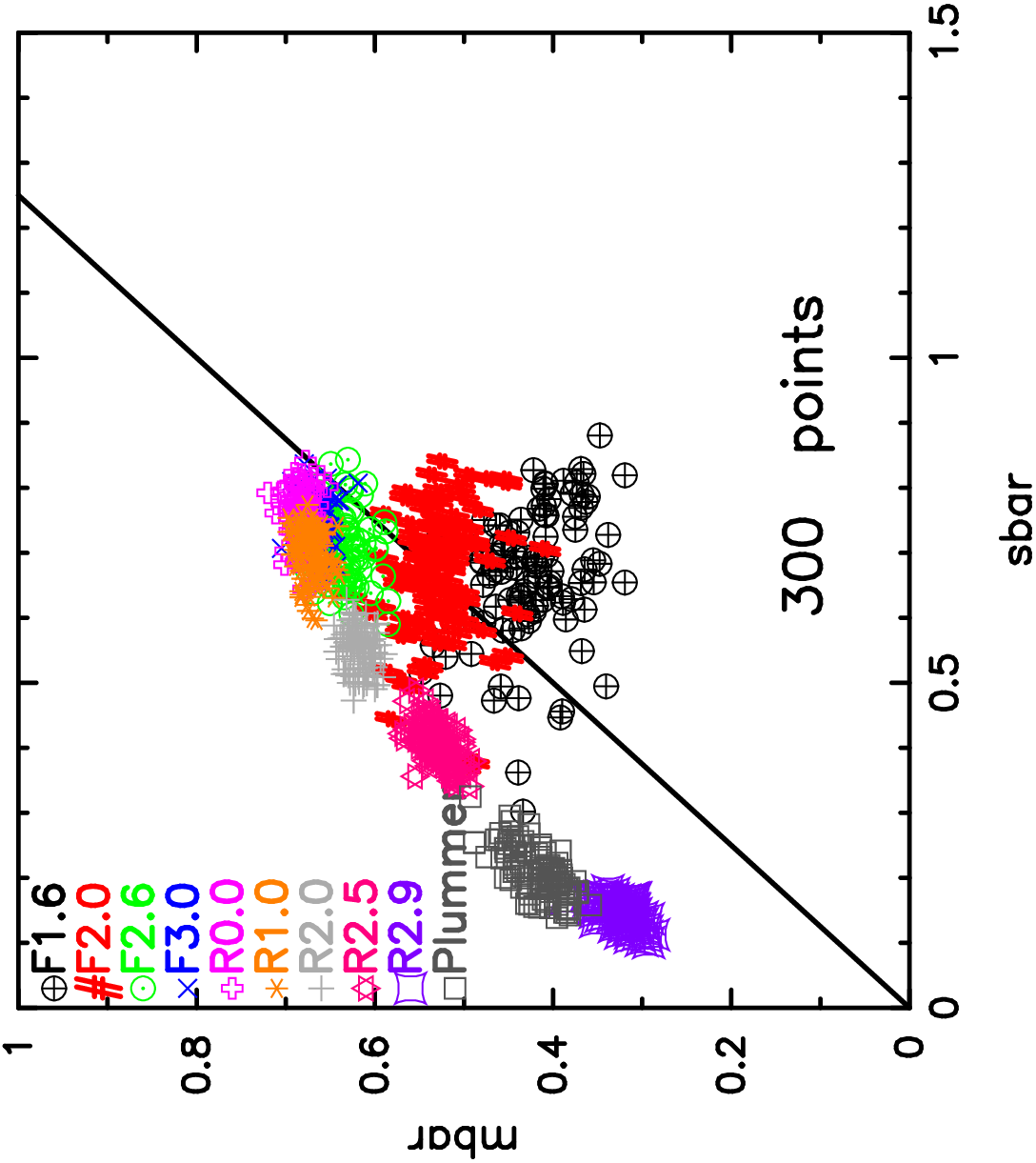}}}
\caption[bf]{The \citet{Cartwright09b} $\bar{m} - \bar{s}$ plot for synthetic star-forming regions containing 300 points. We show the results for ten different geometries, starting with very substructured fractal regions with fractal dimension $D = 1.6$ (the black $\oplus$ symbols) and increasing in fractal dimension (corresponding to increasingly smoother distributions) until the fractals produce a uniform sphere ($D = 3.0$, the blue crosses). We then switch regimes to regions that are smooth and centrally concentrated with a radial density profile $n \propto r^{-\alpha}$, where $\alpha = 0$ indicates a uniform density profile, up to $\alpha = 2.9$ (the purple squashed squares). We also show the results for Plummer spheres (open charcoal squares). The boundary between substructured and smooth distributions using the normalisation technique in \citet{Cartwright04} is shown by the solid black line. We show 100 realisations of each geometry. Panel (a) shows the results where  $\bar{m}$ is normalised to the area of a circle encompassing the outermost point in the distribution and $\bar{s}$ is normalised to the radius of the circle \citep{Cartwright04}. Panel (b) shows results where $\bar{m}$ is normalised to the area of a convex hull, and  $\bar{s}$ is normalised to the radius of a circle with this area \citep{Schmeja06}. Finally, panel (c) shows the results where $\bar{m}$ is normalised to the area of a convex hull, and $\bar{s}$ is normalised to the extent of the outermost point from the mean position of all of the points in the convex hull \citep{Kirk16b}.}
\label{mbar_sbar_test}
  \end{center}
\end{figure*}

The values of the $\mathcal{Q}$-parameter quoted by \citet{Kirk16b} for the three sub regions of Orion~B are all higher than those calculated in Section~\ref{results}. Whereas we calculate low values of $\mathcal{Q}$ which suggest that the cores in the sub-regions follow a substructured distribution, \citet{Kirk16b} find values of $\mathcal{Q}$ that are higher and that appear to be in the regime of $\mathcal{Q}$ that would map to smooth, centrally concentrated distributions.

This discrepancy arises from differences in the methods used to normalise both the mean minimum spanning tree (MST) length $\bar{m}$ and the mean separation between stars, $\bar{s}$. In Fig.~\ref{convex_hull} we show three synthetic star-forming regions, each with a different geometry. Panel (a) of Fig.~\ref{convex_hull} shows a substructured fractal distribution with $D = 1.6$, Panel (b) of Fig.~\ref{convex_hull} shows a uniform fractal with $D = 3.0$ and panel (c) of Fig.~\ref{convex_hull} shows a smooth, centrally concentrated distribution with radial profile ($n \propto r^{-2.9}$). 

In each case, we show the area used to normalise $\bar{m}$ and the radius used to normalise $\bar{s}$ for three different methods. \citet{Cartwright04} normalise their $\mathcal{Q}$-parameter to a circle with area $A$ and radius $R$ (black dashed lines). \citet{Kirk16b} normalise their $\mathcal{Q}$-parameter to a convex hull area $A_{\rm CH}$ and a `radius' equal to the extent of the outermost point of the convex hull from the mean position of the convex hull points, $R_{\rm CH-ex}$ (red solid lines). Finally, \citet{Schmeja06} use this convex hull area $A_{\rm CH}$, but normalise $\bar{s}$ to the radius of a circle, $R_{\rm CH-circ}$ with an area equal to that of the convex hull. 

Irrespective of the geometry of the region, Fig.~\ref{convex_hull} shows that the full convex hull normalisation from \citet{Kirk16b} always produces smaller areas and larger radii than the standard normalisation in \citet{Cartwright04}. This in turn leads to high values of $\mathcal{Q}$ that cannot be mapped to the same scale as the standard normalisation of the $\mathcal{Q}$-parameter. This is demonstrated in Fig.~\ref{qpar_test}, where panel (a) shows the $\mathcal{Q}$-parameter as a function of the number of points in a synthetic distribution. The coloured lines correspond to different geometries, and in panel (a) the lowest (black) line indicates a very substructured distribution, and the sequentially higher lines follow a pattern of decreasing substructure/increasingly smoother and centrally concentrated.

The full convex hull normalisation method suffers from the problem that the normalisation of $\bar{m}$ and $\bar{s}$ for distributions with a low ($<$200) number of points leads to $\mathcal{Q}$ values that do not follow this sequence of regions with the most substructure having lower values of $\mathcal{Q}$. As an example, consider the solid magenta line in panel (c) of Fig.~\ref{qpar_test}, which shows the evolution of the $\mathcal{Q}$-parameter for regions with a smooth distribution and a uniform density profile $n \propto r^{0}$. For regions with fewer than 200 points, the $\mathcal{Q}$-parameter is shown as being lower than a mildly substructured fractal with $D = 2.6$.
 
The normalisation adopted by \citet{Schmeja06} produces almost identical values for $\mathcal{Q}$ to the standard version from \citet{Cartwright04} -- compare panels (a) and (b) in Fig.~\ref{qpar_test}. This is unsurprising as it the radius and area are reduced proportionally (compare the dotted blue line/circle to the dashed black line/circle in Fig.~\ref{convex_hull}).

The \citet{Schmeja06} normalisation does differ from the original \citet{Cartwright04} method in the $\bar{m} - \bar{s}$ plot \citep{Cartwright09b}, which can be used as a further diagnostic check for the amount of substructure present in a region. This method can help distinguish between regimes where the $\mathcal{Q}$-parameter straddles the border between smooth and substructure distributions \citep[e.g.][]{Lomax11,Parker15c}. In Fig.~\ref{mbar_sbar_test} we show the $\bar{m} - \bar{s}$ plot for synthetic star-forming regions containing 300 stars. The boundary between the substructured and smooth regimes for the \citet{Cartwright04} normalisation is shown by the solid line.

The difference between geometries is marginally more distinct in the $\bar{m} - \bar{s}$ plot if we use the \citet{Schmeja06} normalisation (compare panel (b) to panel (a), which is the original \citet{Cartwright04} normalisation). However, the problems with the full convex hull normalisation \citep{Kirk16b} are apparent in panel (c) of Fig.~\ref{mbar_sbar_test}. Different geometries have more overlap in this diagram compared to the \citet{Cartwright04} and \citet{Schmeja06} methods (note the location of several of the black $\oplus$ symbols, which are very substructured fractal distributions, lying in the same parameter space as smooth, very centrally concentrated distributions).  There is also no clear linear boundary between the substructured and smooth regimes (and no obvious alternative location for this boundary).

In summary, the using the full convex hull method to normalise the $\mathcal{Q}$-parameter is flawed, and is the reason behind the spuriously high $\mathcal{Q}$ values quoted in \citet{Kirk16b} for the Orion~B subregions. We recommend using the original normalisation method in \citet{Cartwright04} when calculating the $\mathcal{Q}$-parameter.

\label{lastpage}

\end{document}